\documentclass[twocolumn,showpacs,preprintnumbers,elsart]{revtex4}
\usepackage{makeidx}
\usepackage{amssymb}
\usepackage{amsmath}
\usepackage{mathrsfs}
\usepackage{graphicx}
\usepackage{dcolumn}
\usepackage{bm}
\usepackage[center]{subfigure}
\usepackage{color}
\usepackage{indentfirst}

\begin{document}

\title{Dipolar bright solitons and solitary vortices in a radial lattice}
\author{Chunqing Huang$^{1}$, Lin Lyu$^{1}$, Hao Huang$^{1}$, Zhaopin Chen$
^{2}$,\\ Shenhe Fu$^{3}$, Haishu Tan$^{1}$, Boris A. Malomed$^{2,4,1}$, and Yongyao Li$^{1,5}$}
\email{yongyaoli@gmail.com}
\affiliation
{
$^{1}$School of Physics and Optoelectronic Engineering, Foshan University,
Foshan 528000, China\\
$^{2}$Department of Physical Electronics, School of Electrical Engineering,
Faculty of Engineering, and the Center for Light-Matter Interaction, Tel
Aviv University, Tel Aviv 69978, Israel\\
$^{3}$Department of Optoelectronic Engineering, Jinan University, Guangzhou 510632, China\\
$^{4}$Laboratory of Nonlinear-Optical Informatics, ITMO University, St. Petersburg 197101, Russia\\
$^{5}$College of Electronic Engineering, South China Agricultural University, Guangzhou 510642, China
}

\begin{abstract}
Stabilizing vortex solitons with high values of the topological charge, $S$,
is a challenging issue in optics, studies of Bose-Einstein condensates
(BECs) and other fields. To develop a new approach to the solution of this
problem, we consider a two-dimensional dipolar BEC under the action of an
axisymmetric radially periodic lattice potential, $V(r)\sim \cos (2r+\delta )
$, with dipole moments polarized perpendicular to the system's plane, which
gives rise to isotropic repulsive dipole-dipole interactions (DDIs). Two
radial lattices are considered, with $\delta =0$ and $\pi $, i.e., a
potential maximum or minimum at $r=0$, respectively. Families of vortex gap
soliton (GSs) with $S=1$ and $S\geq 2$, the latter ones often being unstable
in other settings, are completely stable in the present system (at least, up
to $S=11$), being trapped in different annular troughs of the radial
potential. The vortex solitons with different $S$ may stably coexist in
sufficiently far separated troughs. Fundamental GSs, with $S=0$, are found
too. In the case of $\delta =0$, the fundamental solitons are ring-shaped
modes, with a local minimum at $r=0.$At $\delta =\pi $, they place a density
peak at the center.
\end{abstract}

\maketitle



\section{Introduction}

Nonlinear optical and matter waves carrying angular momentum readily
self-trap into vortex modes, which may be considered as two-dimensional dark
solitons supported by a modulationally stable flat background, or bright
solitons with embedded vortices. Experimental and theoretical studies of
vortices is a vast research area in nonlinear optics, studies of
Bose-Einstein condensates (BECs), quantum fluids, and in other fields. The
formation, stability, and dynamics of dark \cite{Pcoullet1989,Swartzlander1992,
Duree1995, DesTorner, Tsubota,Pramana,Allen,Matthews2009,Fetter,Achilleos2012,
Trapani,Liq-cryst,singular} and bright \cite{Dalfovo,Manolo,Berezhiani2001,
Bob,Malomed2002,trapped2,Adhikari2003,Jianke2003,Adhikari2004,quadr-cubic,
bimodal,Neshev2004,Segev,hidden,Zaliznyak,Rotschild2005,Desyatnikov2005,Michinel,
trapped3,Mihalache2006,trapped1,Kartashov2006,Minzoni2007,Jiandong2008,
liangwei2008,Yingji2008,Weiping2009,Skarka2010,Yuwu2013,Min2014,Zhaopin2014,
Kartashov2014,Driben2014,Nir,Dong,liao2017} vortices have been explored
in a great variety of settings, including conservative and dissipative ones,
continuous and discrete media, local and nonlocal interactions, and
different types of the nonlinearity -- cubic (self-focusing and defocusing),
cubic-quintic, saturable, and quadratic (second-harmonic generating). A
recent development has produced unexpected predictions in the form of bright
semi-vortices (bound states of components with vorticities $S=0$ and $S=1$)
in spinor BECs with the spin-orbit coupling\ and contact attractive
interactions, which are stable in free space \cite{SVS1,Yongchang,SVS2,SVS3,SVS4,
Guihua2017},as well as stable gap solitons of the semi-vortex type in the free space
with dipole-dipole interactions (DDIs) \cite{Gap}.

In many cases, bright vortex solitons are stable solely with the unitary
topological charge, $S=1$. In particular, the spin-orbit coupling supports
solitons with vorticities $S=1$ and $S=2$ in its two components, which are
completely unstable \cite{SVS1,Gap}. Vortices in the self-attractive BEC
trapped in an harmonic-oscillator potential also have a stability area
solely for $S=1$ \cite{Dalfovo,trapped2, trapped3,trapped1}. Vortex solitons
in the free space with the cubic-quintic nonlinearity \cite{Manolo} feature
stability regions for $S>1$, but they are very narrow, starting from $S=3$
\cite{Bob}. Typically, the vortex solitons with $S>1$ are subject to
azimuthal perturbations which break them into $S$ fragments \cite%
{Soto-Crespo1991,Skrybin1998,trapped2,trapped1}. This instability has been
demonstrated experimentally for vortex beams propagating in saturable
self-focusing media \cite{Soljacic1998, Desyantikov2001, Soljaic2001}, as well
as in quadratic ones \cite{Dima}.

Sufficiently large stability regions in the parametric space for vortex
solitons with $S>1$ were found in axisymmetric potential lattices with the
Bessel functional profile in the radial direction, combined with the
self-defocusing nonlinearity \cite{Kartashov2005Bessel}. In that case, a
deeper lattice is required to stabilize solitary vortices with higher values
of $S$. Because the Bessel potential vanishes at $r\rightarrow \infty $, the
total norm of modes trapped in it under the action of self-defocusing,
strictly speaking, diverges in the infinite space. Truly confined gap
solitons (GSs), i.e., solitons whose chemical potential falls in one of bandgaps
generated by the underlying potential lattice, were constructed considering
the combination of the self-defocusing cubic nonlinearity
and a radially-periodic potential, $\sim
\cos \left( 2kr\right) $, where $r$ is the radial coordinate \cite%
{Bakhtiyor2006}. However, only radial GSs with $S=0$ were found to be
completely stable in the latter model, while all confined vortices featured
a weak azimuthal instability. Self-trapped vortices, which remain stable, at
least, up to $S=5$, were recently found in a model of a polariton type,
which combines the self-repulsive contact nonlinearity of a two-component
BEC and effective nonlocal self-attraction mediated by the microwave field
generated by transitions between two components resonantly coupled by the
field \cite{Dong}. Nonlocal interactions, considered in Ref.
\cite{Dong}, or in the present work (see below), introducing their own radial scale,
provide more options in the interplay with the radially-periodic
lattice, which helps, in particular, to stabilize vortex GS modes against
the azimuthal instability.

The objective of the present work is to predict stable GSs with $S=0$ and $%
S\geq 1$ in a dipolar BECs trapped in a radially periodic potential, with
dipole moments polarized perpendicular to the system's plane, which gives
rise to the isotropic repulsive DDI. In earlier works, DDIs were used to
predict stable one-dimensional \cite{Santos11,epsilon01,Lauro,Bland2015} and
two-dimensional \cite{Pedri2005,Tikhonenkov2008,Tikhonenkov2208,Gligoric1,
Gligoric2,koberle2012,Yongyao2013,Raghuandan2015,Xuyong2015,Huaiyu2006,
Muruganandam2011} solitons in other settings. In addition, it was found that
quadrupole-quadrupole interactions are also able to create stable two-dimensional
solitons \cite{Q,Jiasheng}.However, the free-space DDI per se cannot stabilize
vortex solitons with $S>1 $ \cite{Tikhonenkov2208}. In this work, we demonstrate
that ring-shaped vortex GSs with higher values of $S$ (at least, up to $S=11$)
are readily made stable by the combined effect of the radial lattice potential and
repulsive isotropic DDIs. Furthermore, double and multiple sets of
concentric vortex solitons, with different topological charges, may stably
coexist, if placed in different annular potential troughs of the radial
lattice. In that case, vorticity jumps take place at zero-amplitude notches
separating the concentric vortices. The latter property was not reported in
previously considered two-dimensional models.

The paper is structured as follows. The model is introduced in Sec. II,
which is followed by presentation of numerical results for the fundamental ($%
S=0$) and vortex ($S\geq 1$) GSs in Sec. III. The stability of the solitons
is verified by means of systematic direct simulations. The paper is
concluded by Sec. IV.

\section{The model}

\begin{figure}[t]
\subfigure[]{\includegraphics[width=0.49\columnwidth]{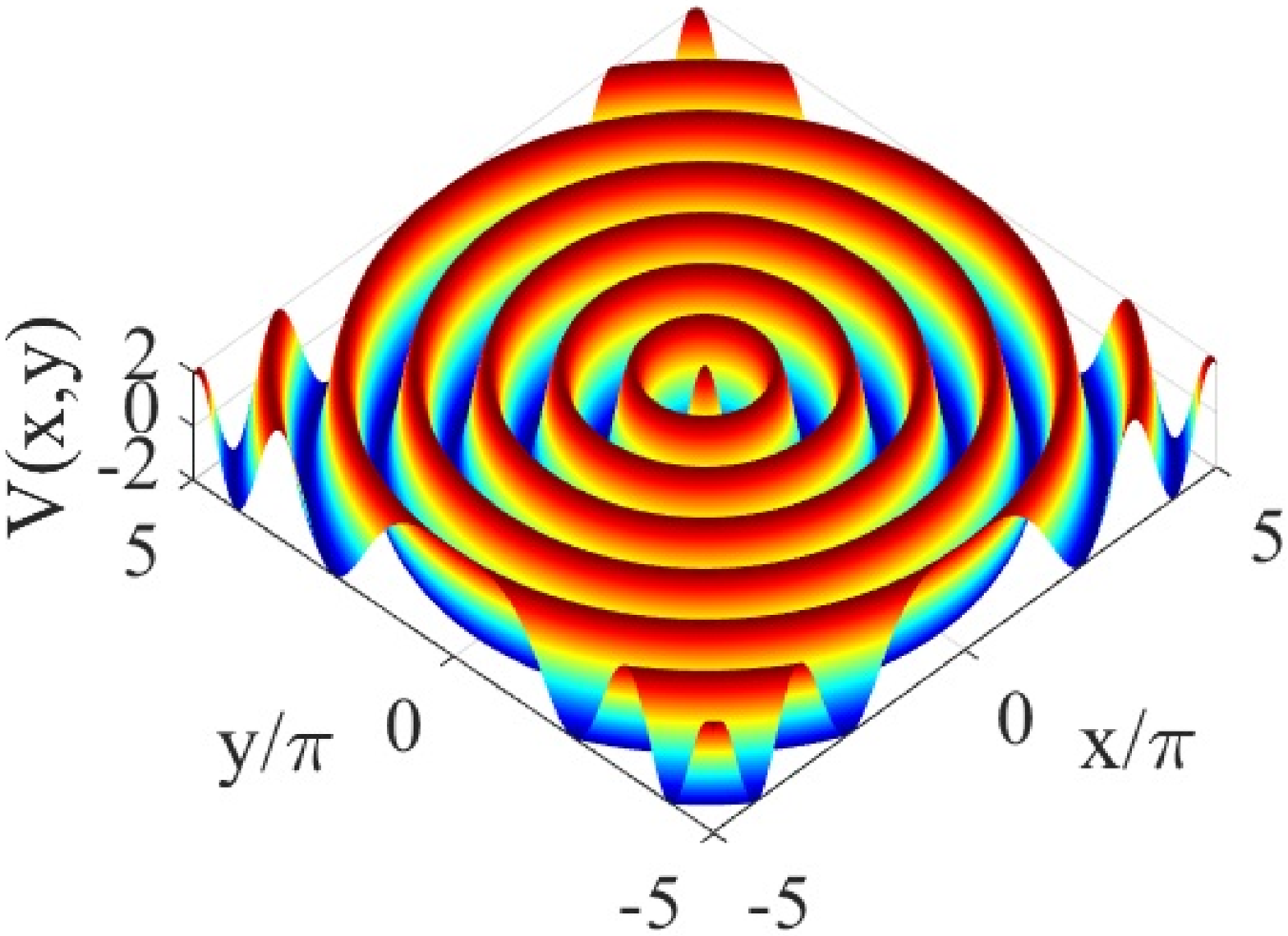}} %
\subfigure[]{\includegraphics[width=0.49\columnwidth]{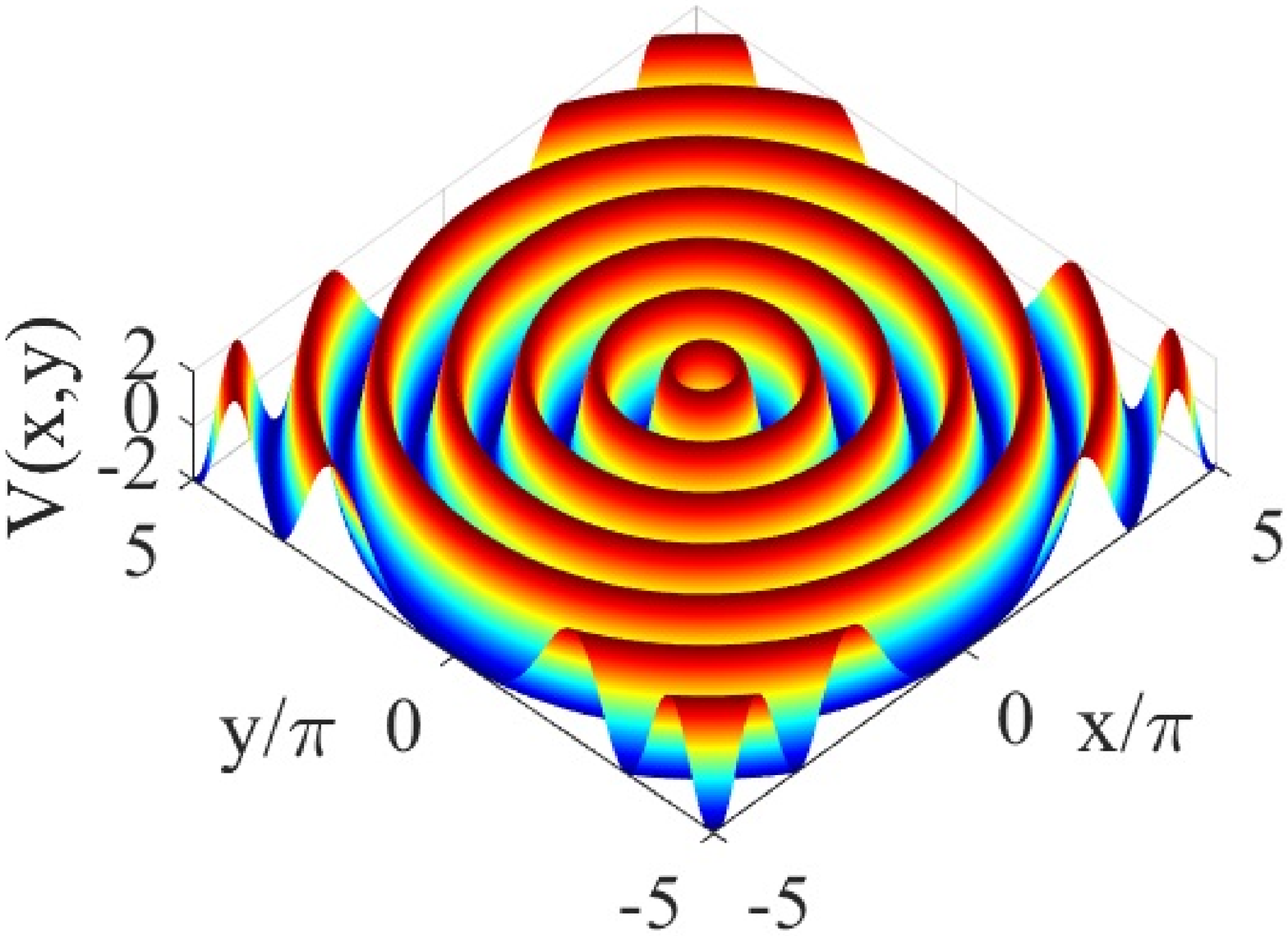}}
\caption{Radial potentials (\protect\ref{ring}) with $V_{0}=1$, for $\protect%
\delta =0$ (a) and $\protect\delta =\protect\pi $ (b).}
\label{potentialwell}
\end{figure}

According to what is said above, we consider an effectively two-dimensional
setting, modeled by Gross-Pitaevskii equation, which is written in the
scaled form:
\begin{equation}
\begin{aligned} i{\frac{\partial }{\partial t}}\Psi
(\mathbf{r},t)=-{\frac{1}{2}}\nabla ^{2}\Psi (\mathbf{r},t)+V(r)\Psi
(\mathbf{r},t)\\ +\kappa \Psi (\mathbf{r},t)\int
R(\mathbf{r}-\mathbf{r^{\prime }})|\Psi (\mathbf{r^{\prime
}},t)|^{2}d\mathbf{r^{\prime }}, \label{fulleq} \end{aligned}
\end{equation}%
where $\mathbf{r}=\left\{ x,y\right\} $ is the set of coordinates, $\nabla
^{2}=\partial _{x}^{2}+\partial _{y}^{2}$ is the respective Laplacian, $\Psi
(\mathbf{r},t)$ is the mean-field wave function, and $\kappa >0$ is the
strength of the DDI, with the isotropic kernel corresponding to the
particles' dipolar moments polarized perpendicular to the $\left( x,y\right)
$ plane:
\begin{equation}
R(\mathbf{r}-\mathbf{r^{\prime }})={\frac{1}{[\epsilon ^{2}+(\mathbf{r}-%
\mathbf{r^{\prime }})^{2}]^{3/2}}}.  \label{rkerneleq}
\end{equation}%
Here, cutoff $\epsilon $ is the regularization parameter, which is
determined by the confinement of the three-dimensional condensate in the
transverse direction \cite{Santos11,epsilon01}. Further, the axisymmetric
radially-periodic lattice potential is taken as
\begin{equation}
V(r)=V_{0}\mathrm{cos}(2r+\delta ),  \label{ring}
\end{equation}%
where $r=\sqrt{x^{2}+y^{2}}$, the depth of the lattice potential is $2V_{0}>0
$, the radial period is fixed to be $\pi $ by scaling, and $\delta $ is a
phase constant. Here, we focus on the consideration of two most essential
cases, \textit{viz}., $\delta =0$ and $\delta =\pi $, which correspond to a
potential maximum or minimum at the center, $r=0$, respectively, see Fig. %
\ref{potentialwell}.
We look for stationary axisymmetric states with\ chemical potential $\mu $
and integer vorticity $S$ as solutions to Eq. (\ref{fulleq}) in the form of
\begin{equation}
\Psi (\mathbf{r},t)=\psi (r)\exp \left( iS\theta -i\mu t\right) ,
\label{solution}
\end{equation}%
where $\theta $ is the angular coordinate. Self-trapped GS solutions are
characterized by the total norm,
\begin{equation}
N=2\pi \int_{0}^{\infty }{\psi }^{2}{(r){rdr}},  \label{ppeq}
\end{equation}%
and the angular momentum, $M=SN$. Its energy is
\begin{equation}
E=E_{\mathrm{K}}+E_{\mathrm{V}}+E_{\mathrm{DDI}},  \label{E}
\end{equation}%
where $E_{\mathrm{K}}$, $E_{\mathrm{V}}$ and $E_{\mathrm{DDI}}$ are the
kinetic, potential, and DDI terms, respectively:

\begin{equation}
E_{\mathrm{K}}=\int \left\vert \nabla \psi \right\vert ^{2}d\mathbf{r} \\
\equiv {\pi }\int_{0}^{\infty }\left[ \left( \frac{d\psi }{dr}\right) ^{2}+%
\frac{S^{2}}{r^{2}}\psi ^{2}(r)\right] rdr,  \label{K}
\end{equation}

\begin{eqnarray}
&&E_{\mathrm{V}}=2\pi \int {V(}r{)\psi }^{2}{(r)rd}r,  \label{V} \\
&&E_{\mathrm{DDI}}={\frac{\kappa }{2}}\iint {R(\mathbf{r}-\mathbf{r}^{\prime
})|\psi (\mathbf{r})|^{2}{|\psi (\mathbf{r}^{\prime })|}^{2}}d\mathbf{r}d%
\mathbf{r}^{\prime }.  \label{EDDI}
\end{eqnarray}

Two-dimensional bright GSs can be supported by the interplay of the radially
periodic potential and repulsive interaction \cite{Bakhtiyor2006}. In this
work, we focus on the DDI, which was not previously considered in the
present setting, neglecting contact interactions, which can be effectively
suppressed by means of the Feshbach resonance \cite{Feshbach}. In fact,
effects of adding moderately strong contact interactions to the DDI were
checked too (not shown in detail in this paper, as no dramatic changes
in the results were observed in that case).

Numerical simulations have been carried out by dint of algorithm of PCSOM
\cite{YangJK}, fixing $\kappa \equiv 1$ by means of scaling, and, typically,
taking $\epsilon =0.5$, which is small enough in comparison with the potential's
period, $\pi $, making it possible to produce generic results. It was additionally
checked that taking still smaller $\epsilon$ (e.g., $0.25$) does not produce any
conspicuous change in the results. The stability
of stationary soliton solutions was tested by means of real-time
propagation, which was implemented with the help of the standard split-step
-- fast-Fourier-transform algorithm.

\section{Numerical results}

\subsection{The radial lattice with $\protect\delta =0$ (potential maximum
at the center)}

\begin{figure*}[t]
\subfigure[]{\includegraphics[width=0.55\columnwidth]{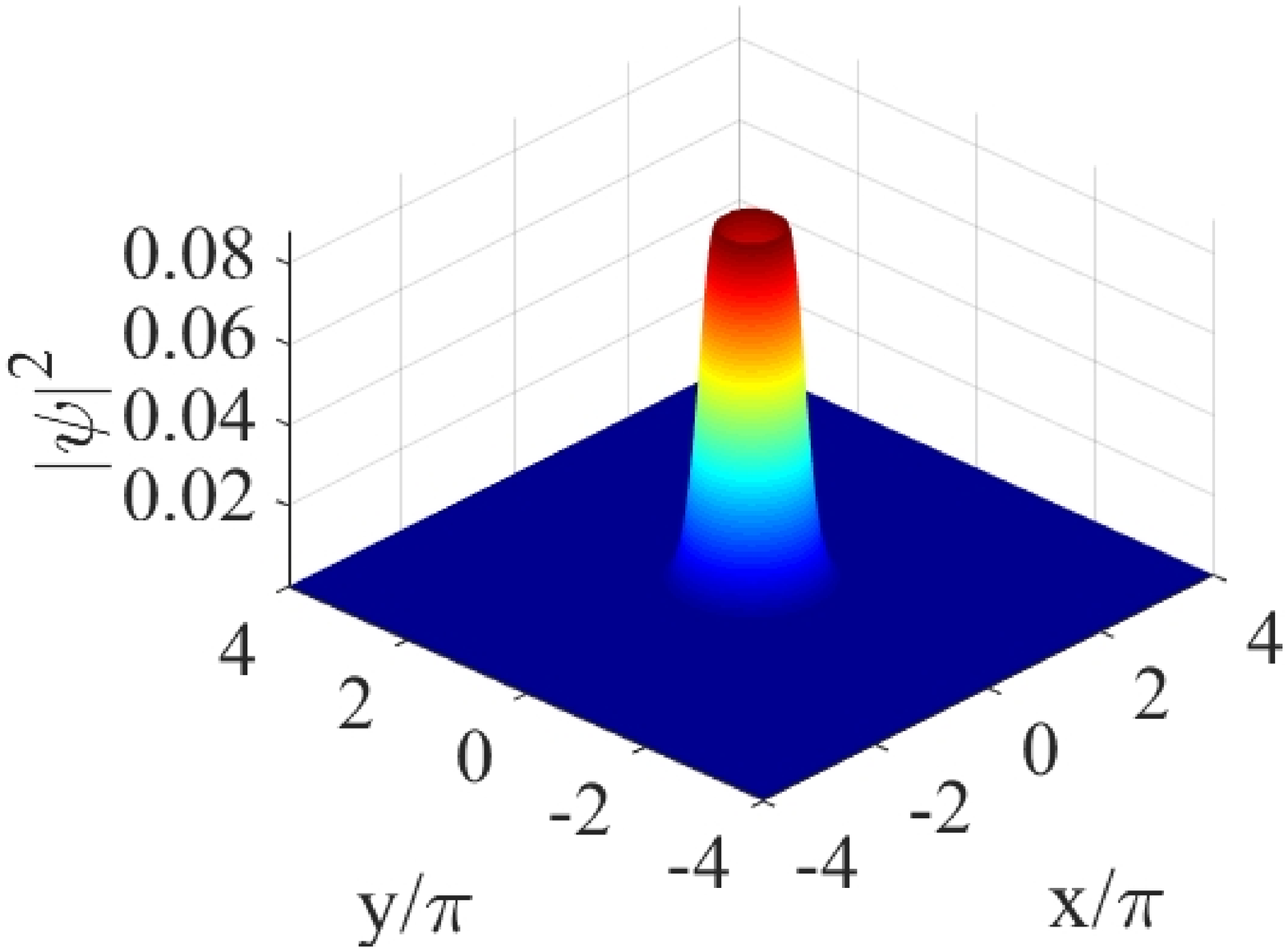}} %
\subfigure[]{\includegraphics[width=0.55\columnwidth]{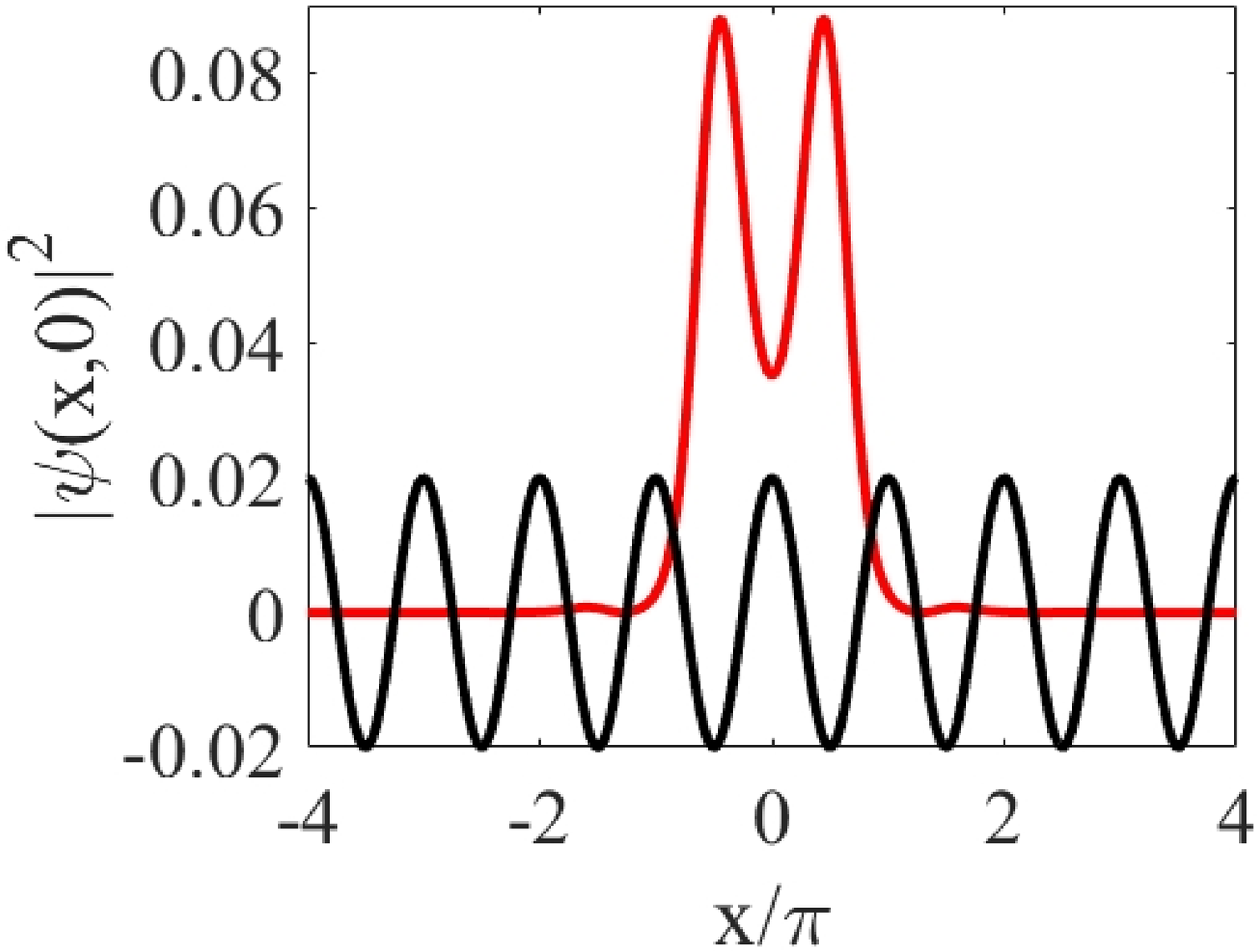}} %
\subfigure[]{\includegraphics[width=0.55\columnwidth]{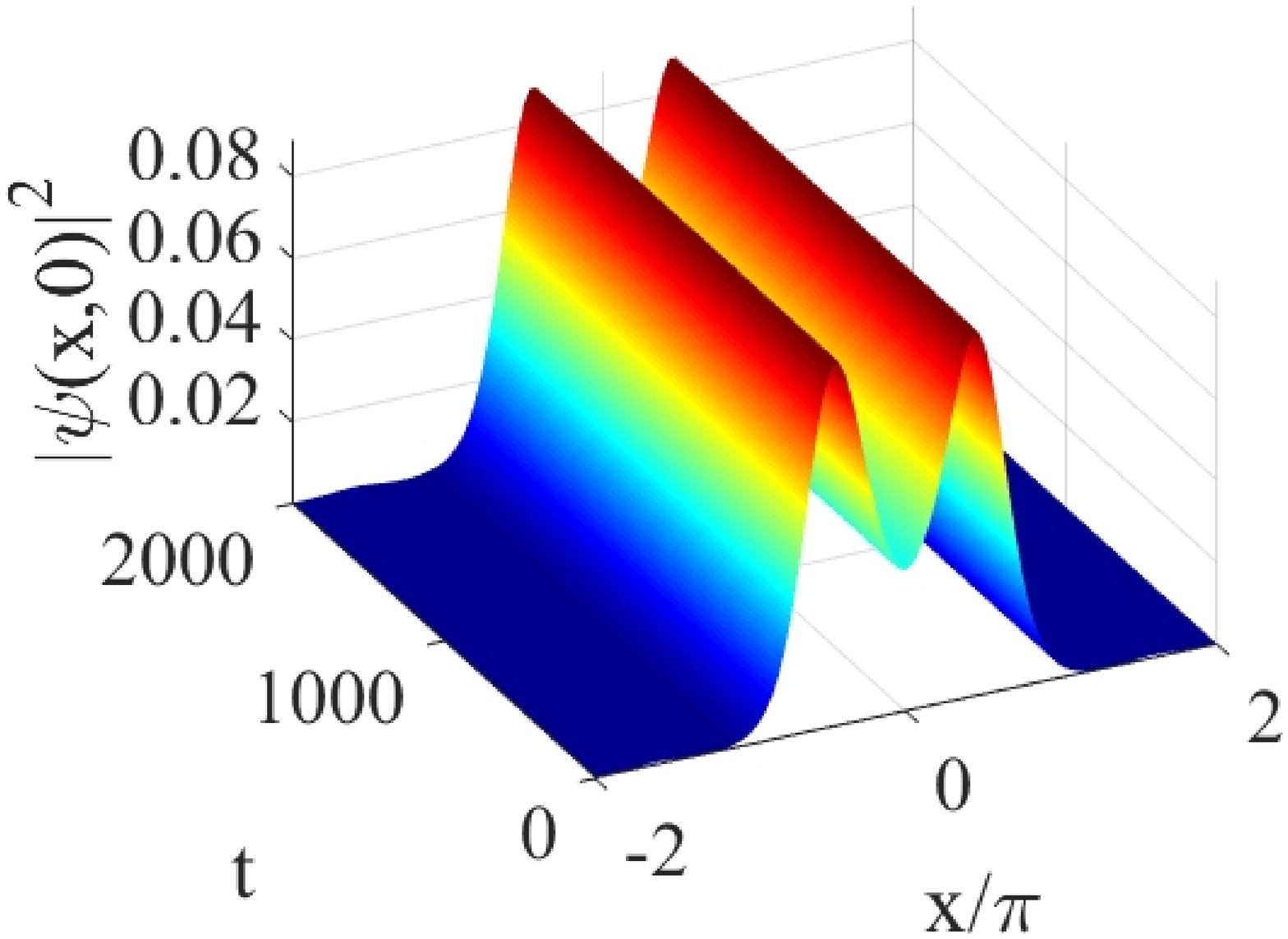}}
\caption{(Color online) A typical fundamental ($S=0,n=1$) ring-shaped gap
soliton for $\protect\delta =0$, other parameters being $N=1.3$ and $V_{0}=1$%
. (a) The density of stationary wave function in the $(x,y)$ plane. (b) Its
cross section, $|\protect\psi \left( x,0\right) |^{2}$, along $y=0$. (c) The
cross-section of the real-time simulation, which demonstrates stability of
the soliton.}
\label{Ringsoliton}
\end{figure*}

\begin{figure*}[t]
\subfigure[]{\includegraphics[width=0.55\columnwidth]{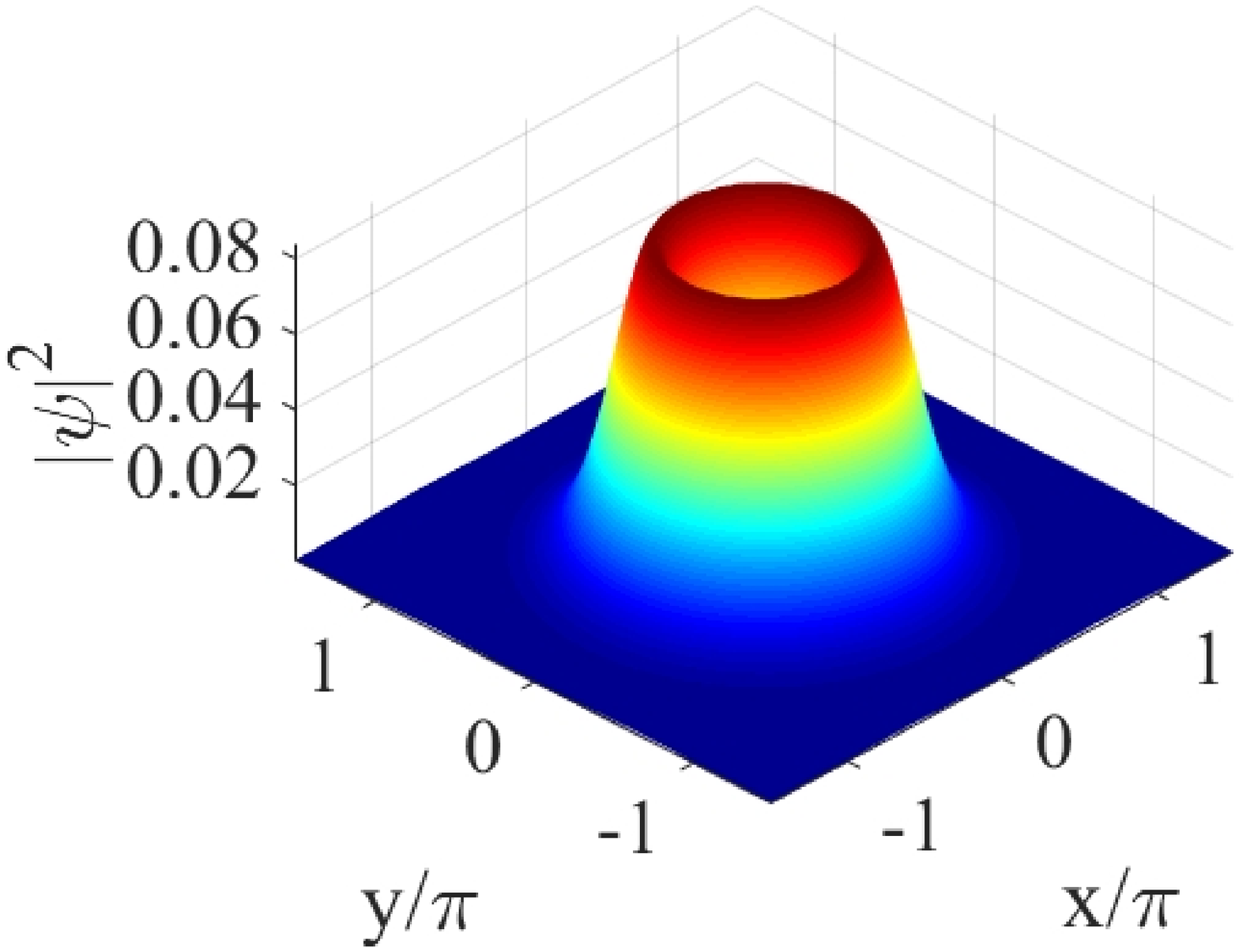}} %
\subfigure[]{\includegraphics[width=0.55\columnwidth]{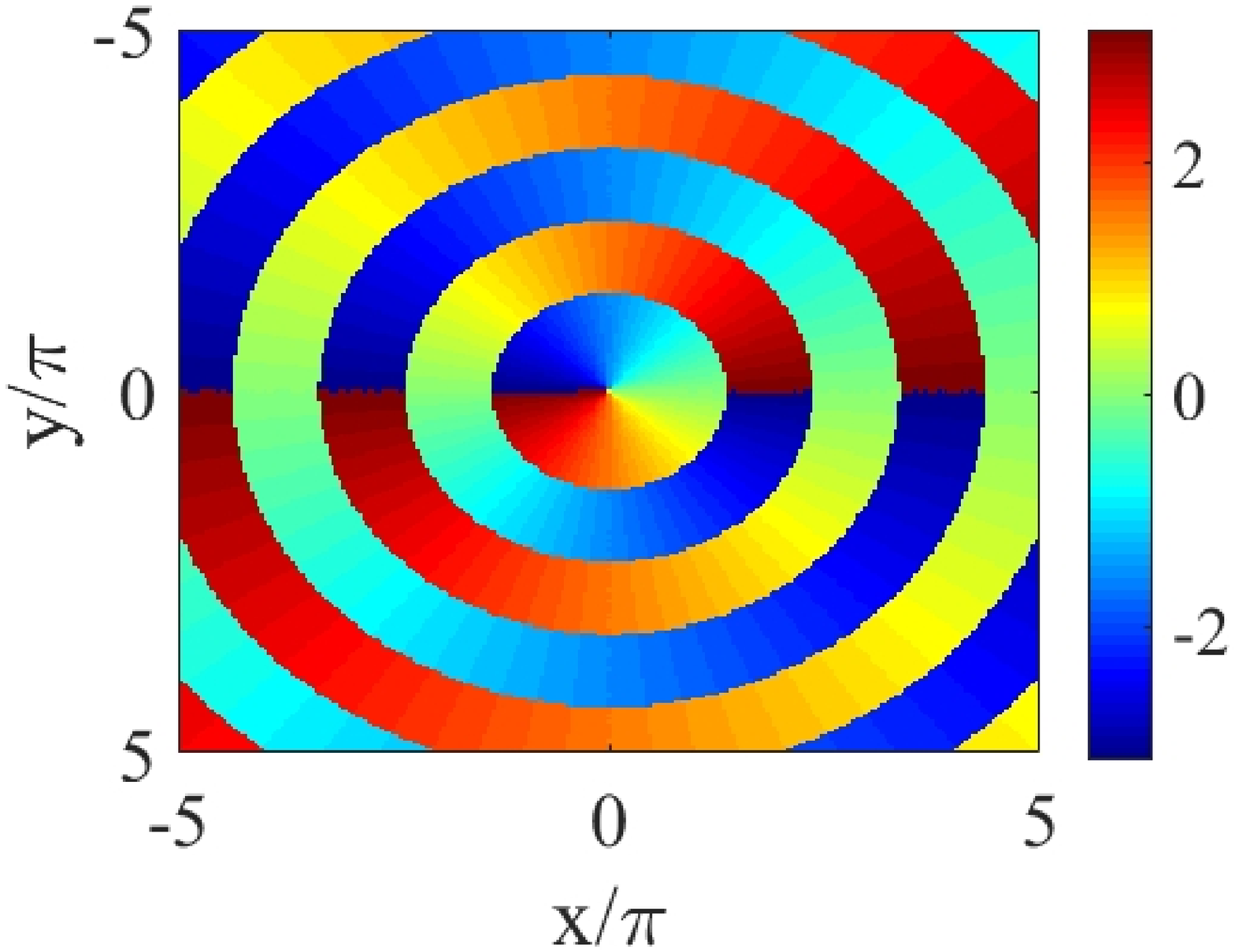}} %
\subfigure[]{\includegraphics[width=0.55\columnwidth]{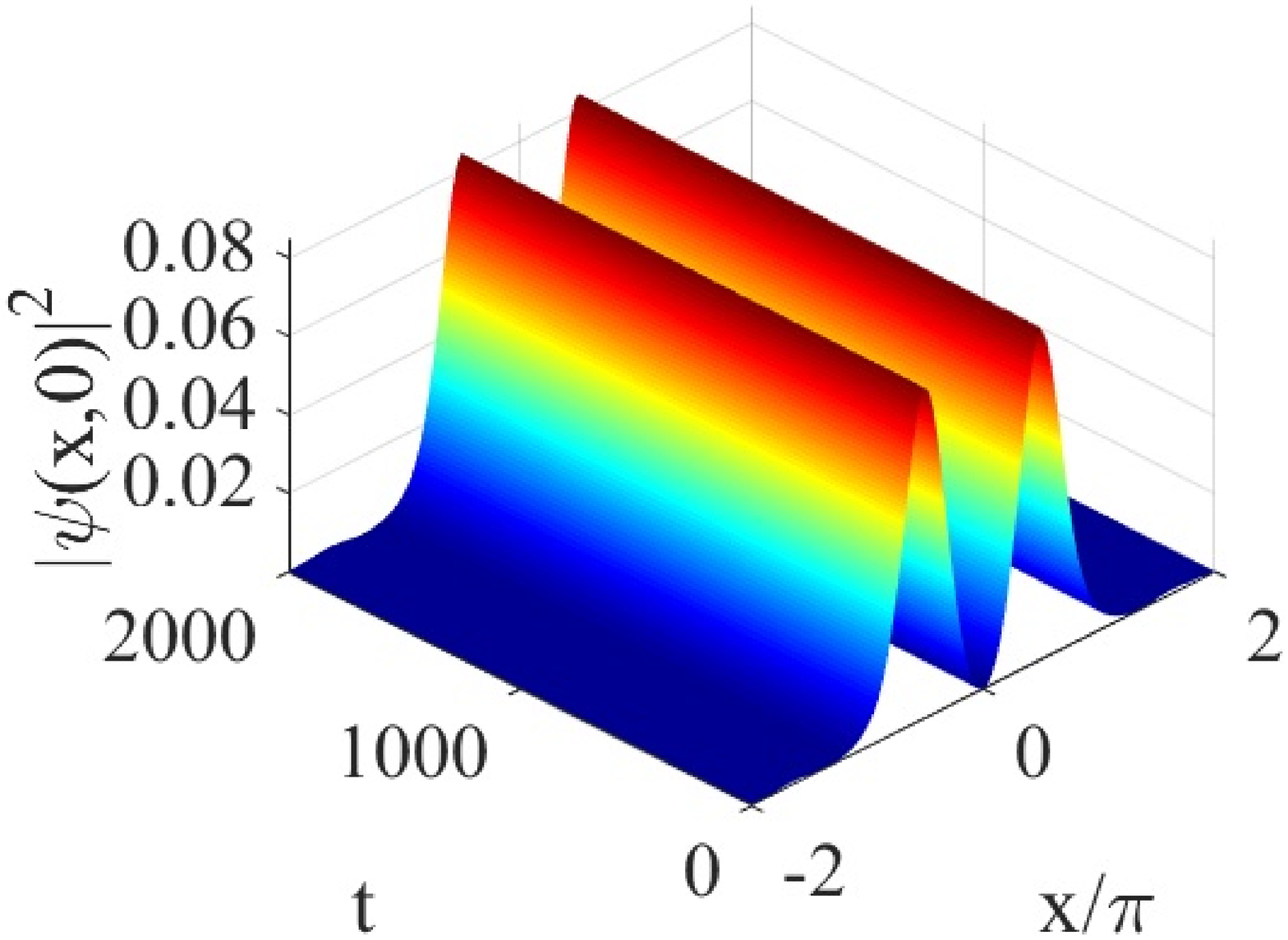}} %
\subfigure[]{\includegraphics[width=0.55\columnwidth]{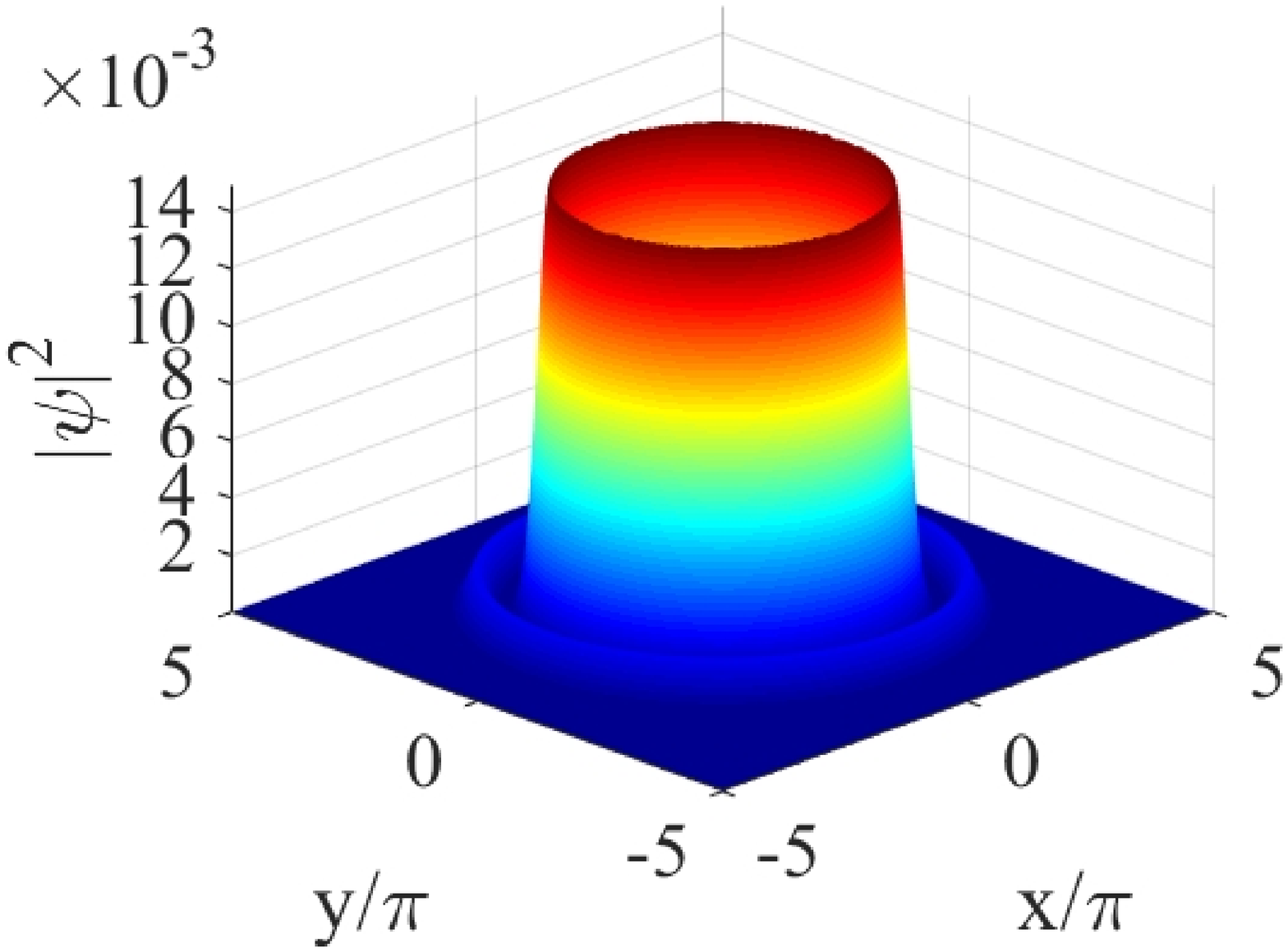}} %
\subfigure[]{\includegraphics[width=0.55\columnwidth]{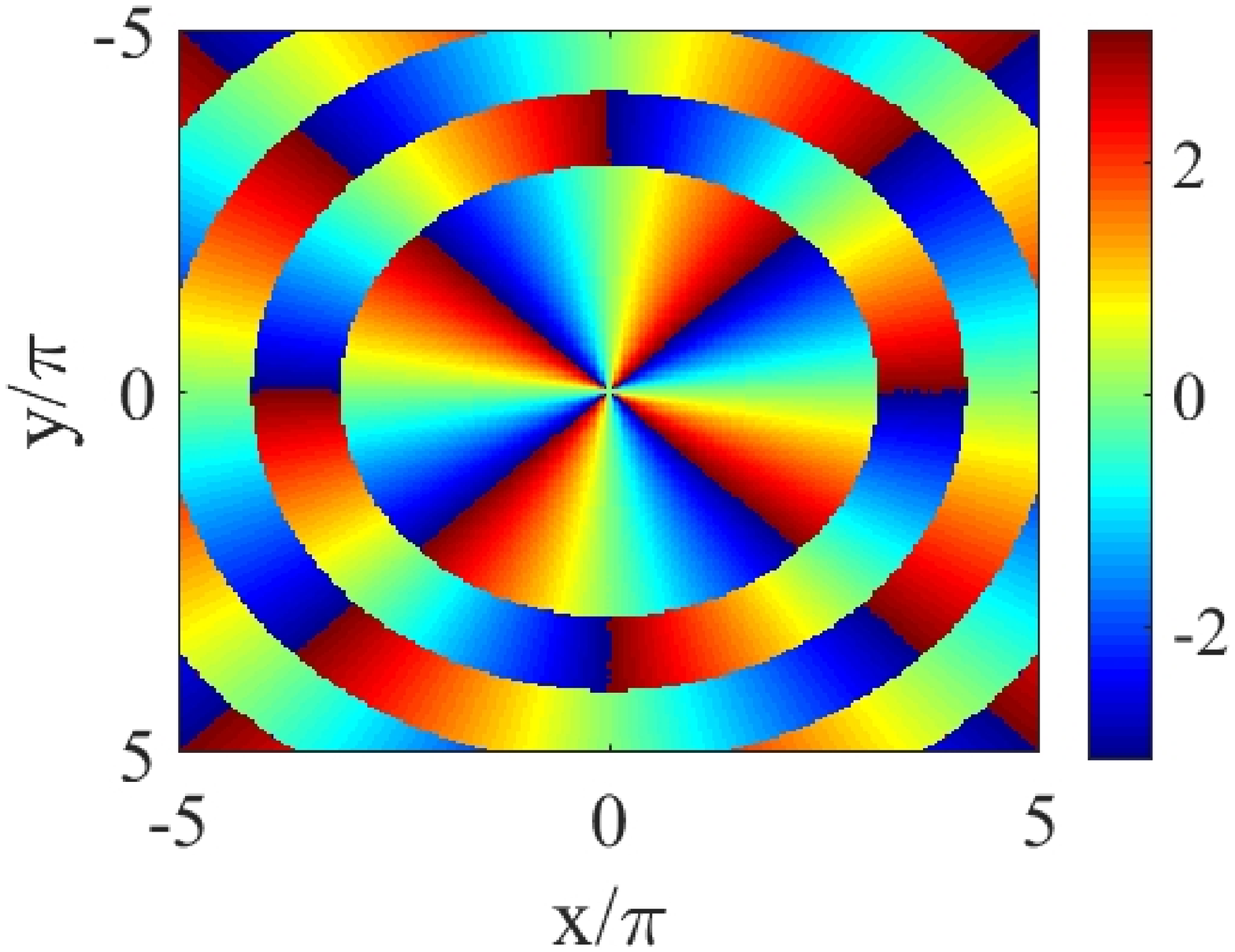}} %
\subfigure[]{\includegraphics[width=0.55\columnwidth]{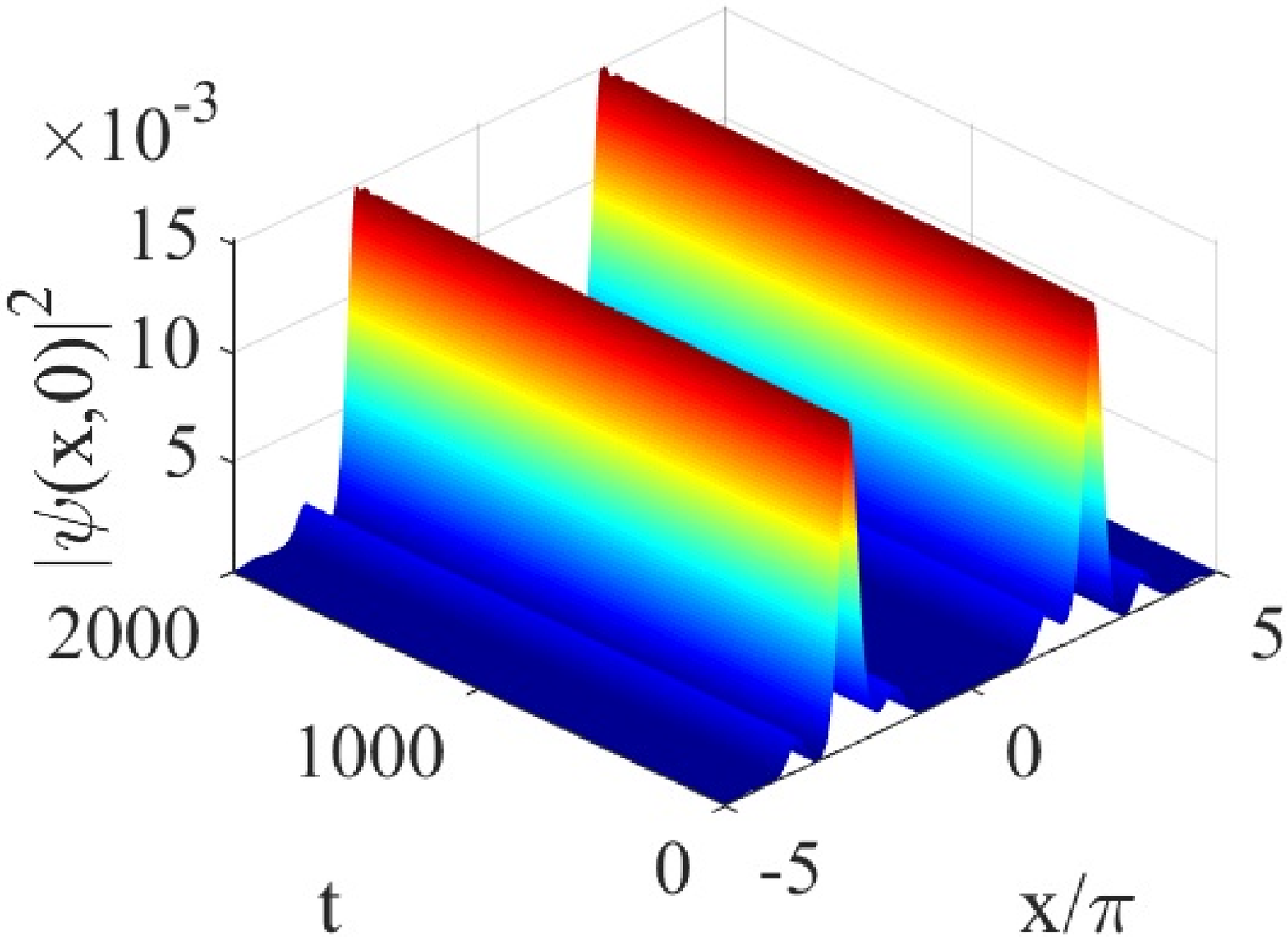}}
\caption{(Color online) Two typical examples of stable gap-vortex solitons
for $\protect\delta =0$. (a) and (b): Density and phase profiles of the
vortex soliton for $S=1,n=1$ [$n$ is defined as per Eq. (\protect\ref{min}%
)]. (c) The cross-section of the real-time evolution, corroborating the
stability of the vortex soliton. (d,e,f) The same as in (a,b,c),
respectively, but for $S=4,n=3$. The norm of both vortex solitons is $N=1.3$%
, and the strength of the radial potential is $V_{0}=1$.}
\label{Vortexsoliton}
\end{figure*}

When radial potential (\ref{ring}), has a maximum at the center, Eq. (\ref%
{fulleq}), naturally, cannot produce a fundamental soliton ($S=0$) with a
density peak at $r=0$. Instead, the model readily produces stable GSs with $%
S=0,1,2,3,4,..$. which feature a density minimum at the center, and the main
radial density peak trapped in a trough with the bottom at one of potential
minima,%
\begin{equation}
r_{\min }=\pi \left( n-\frac{1}{2}\right) ,~n=1,2,3,4,...,  \label{min}
\end{equation}%
where $n$ is the number of the radial minimum. Typical examples of such
ring-shaped solitons are displayed in Fig. \ref{Ringsoliton} for $S=0$, and
in Fig. \ref{Vortexsoliton} for $S=1$ and $S=4$. The GSs with $S=0$ and $1$
place their density maxima close to $n=1$, while the vortex with $S=4$
chooses $n=3$. It is worthy to note than the latter vortex mode, with a high
value of the topological charge, $S=4$, displayed in Fig. \ref{Vortexsoliton}%
(d-f), is definitely stable, on the contrary to a majority of models where
it would be unstable. In fact, we have obtained stable vortex solitons with
topological charges up to $S=11$, as indicated below in Figs. \ref{muVortex}%
(a) and \ref{deltapiVortex}(d).

\begin{figure*}[t]
\subfigure[]{\includegraphics[width=0.55\columnwidth]{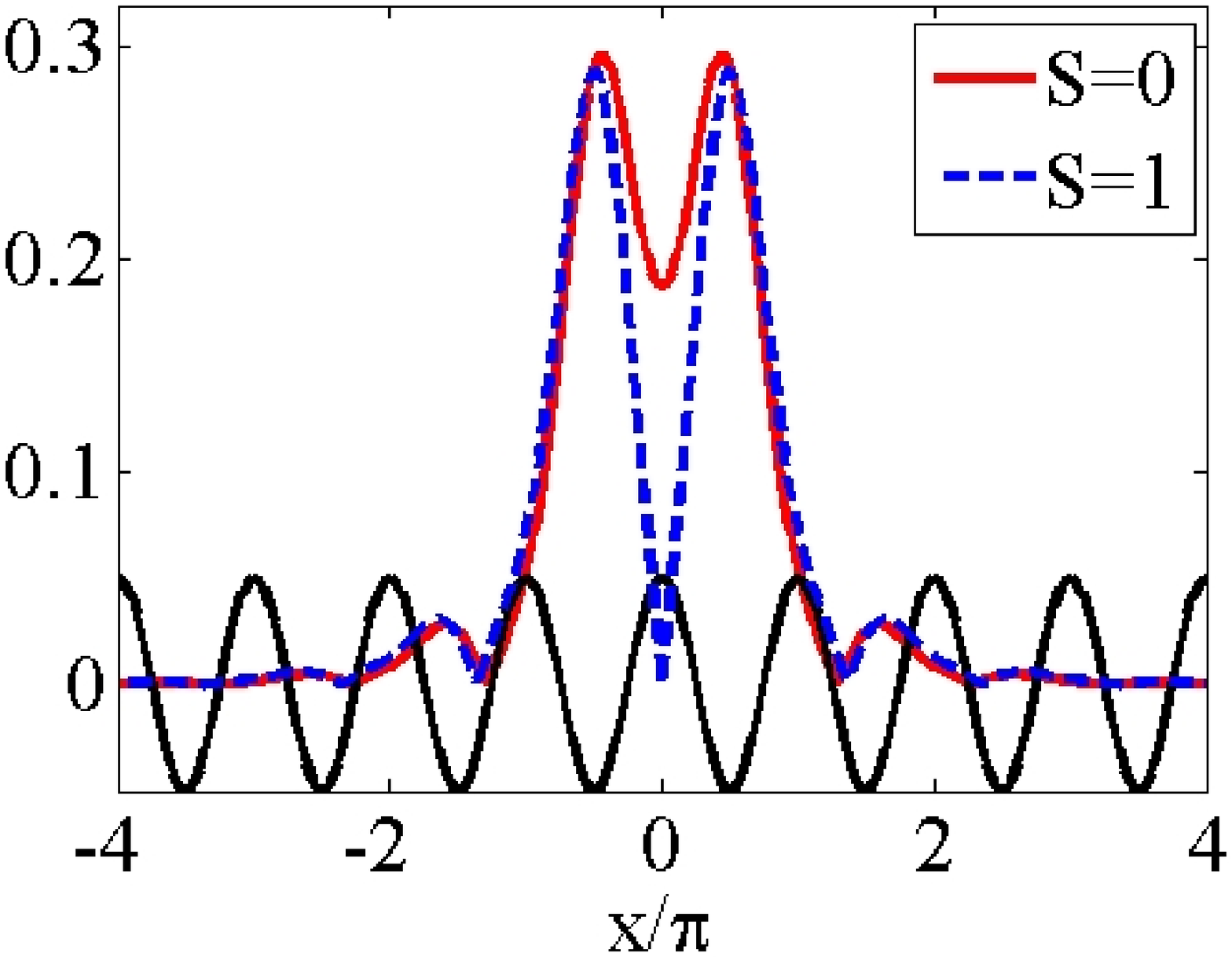}} %
\subfigure[]{\includegraphics[width=0.55\columnwidth]{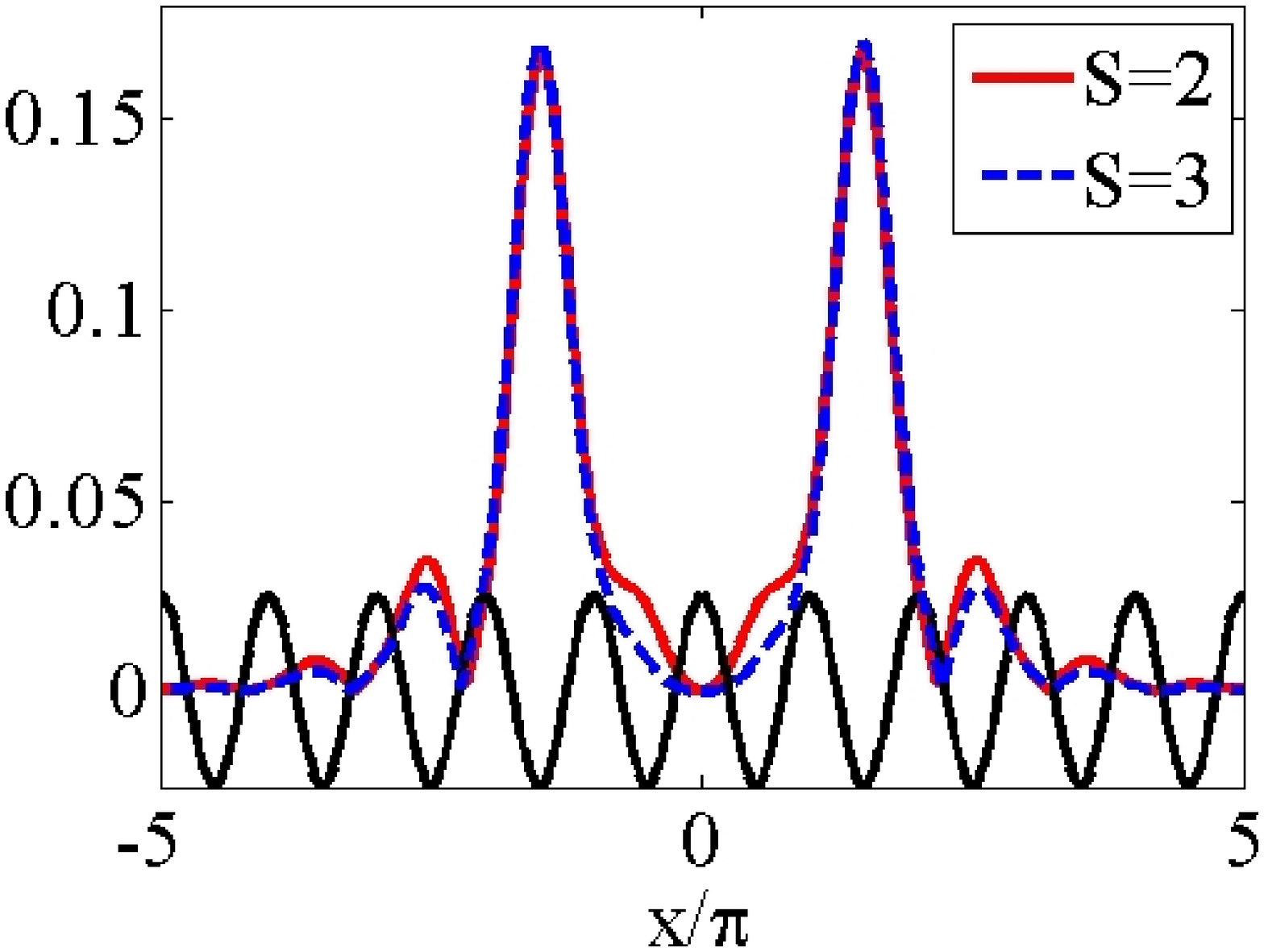}} %
\subfigure[]{\includegraphics[width=0.55\columnwidth]{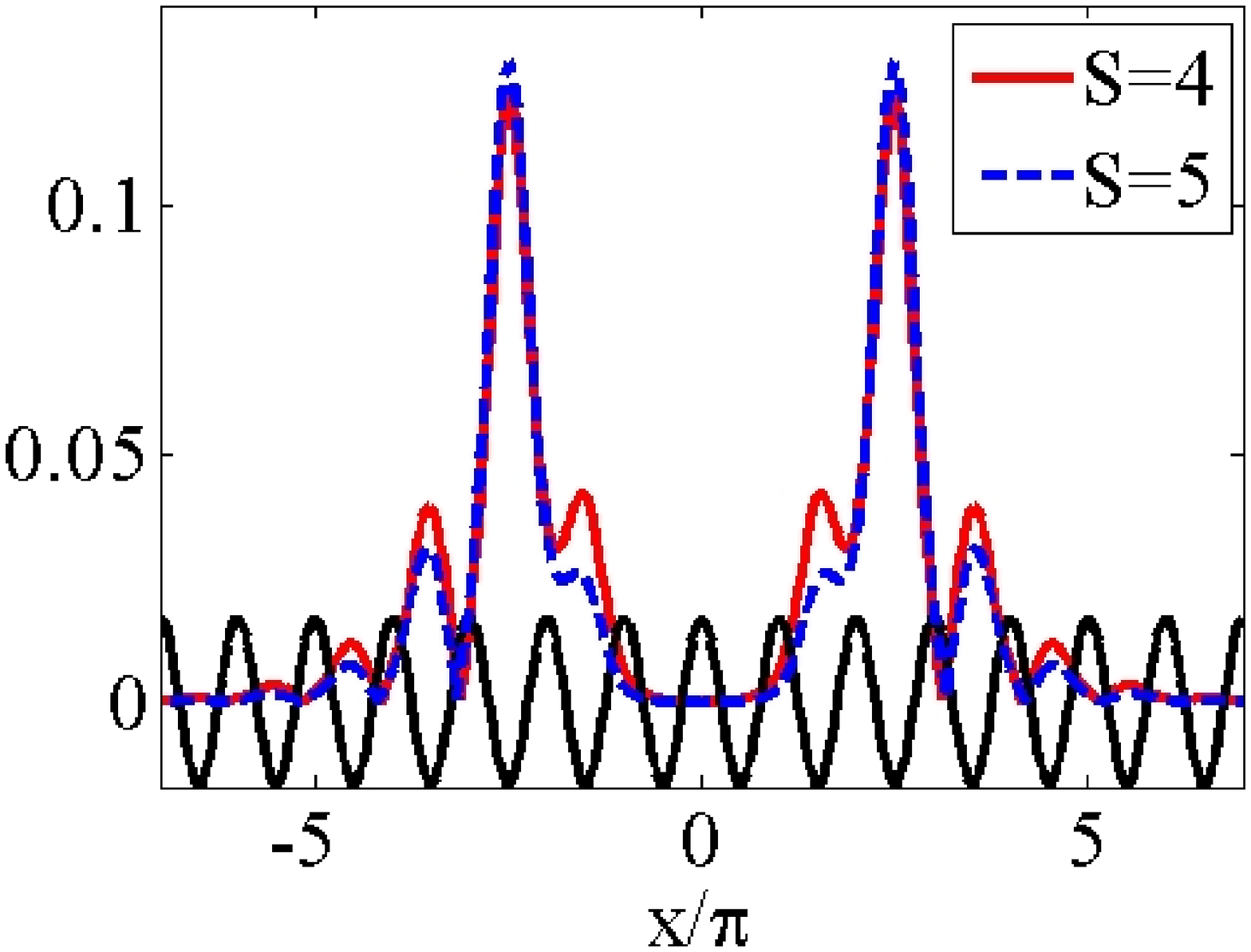}}
\caption{(Color online) (a) The cross-section $|\protect\psi (x,0)|$ of
profiles of the gap solitons trapped in different annular potential troughs
(here we display the absolute value of the field, rather than the density,
to display the profiles in a clearer form). (a) The solitons with $S=0$ and $%
1$ in the first trough, which corresponds to $n=1$ in Eq. (\protect\ref{min}%
). (b) $S=2$ and $3$, in the trough with $n=2$. (c) $S=4$ and $5$, in the
trough with $n=3$. The norm of all the gap solitons displayed here is $N=1.3$%
, and the half-depth of the periodic potential is $V_{0}=1$.}
\label{abspsi}
\end{figure*}

\begin{figure}[t]
\subfigure[]{\includegraphics[width=0.49\columnwidth]{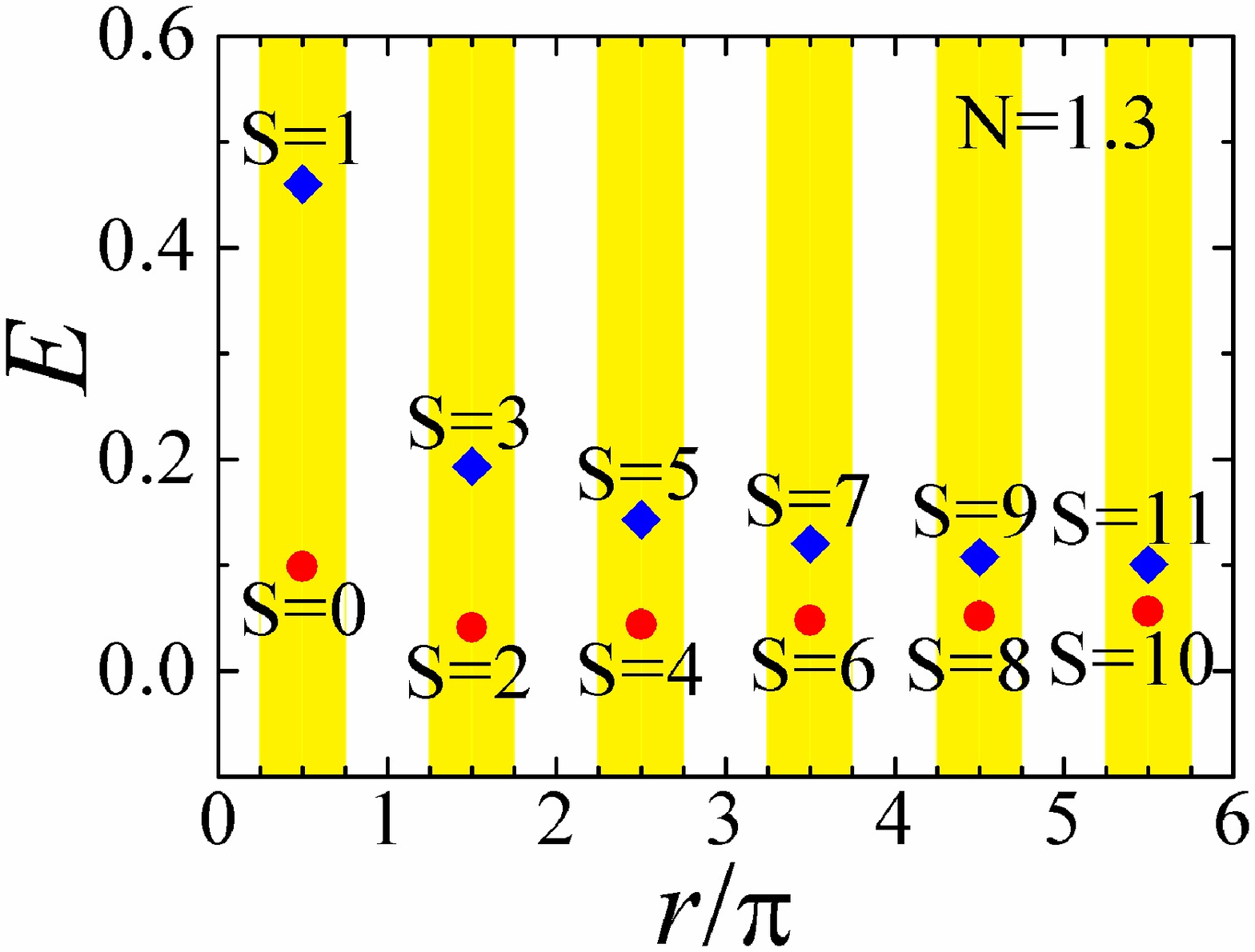}} %
\subfigure[]{\includegraphics[width=0.49\columnwidth]{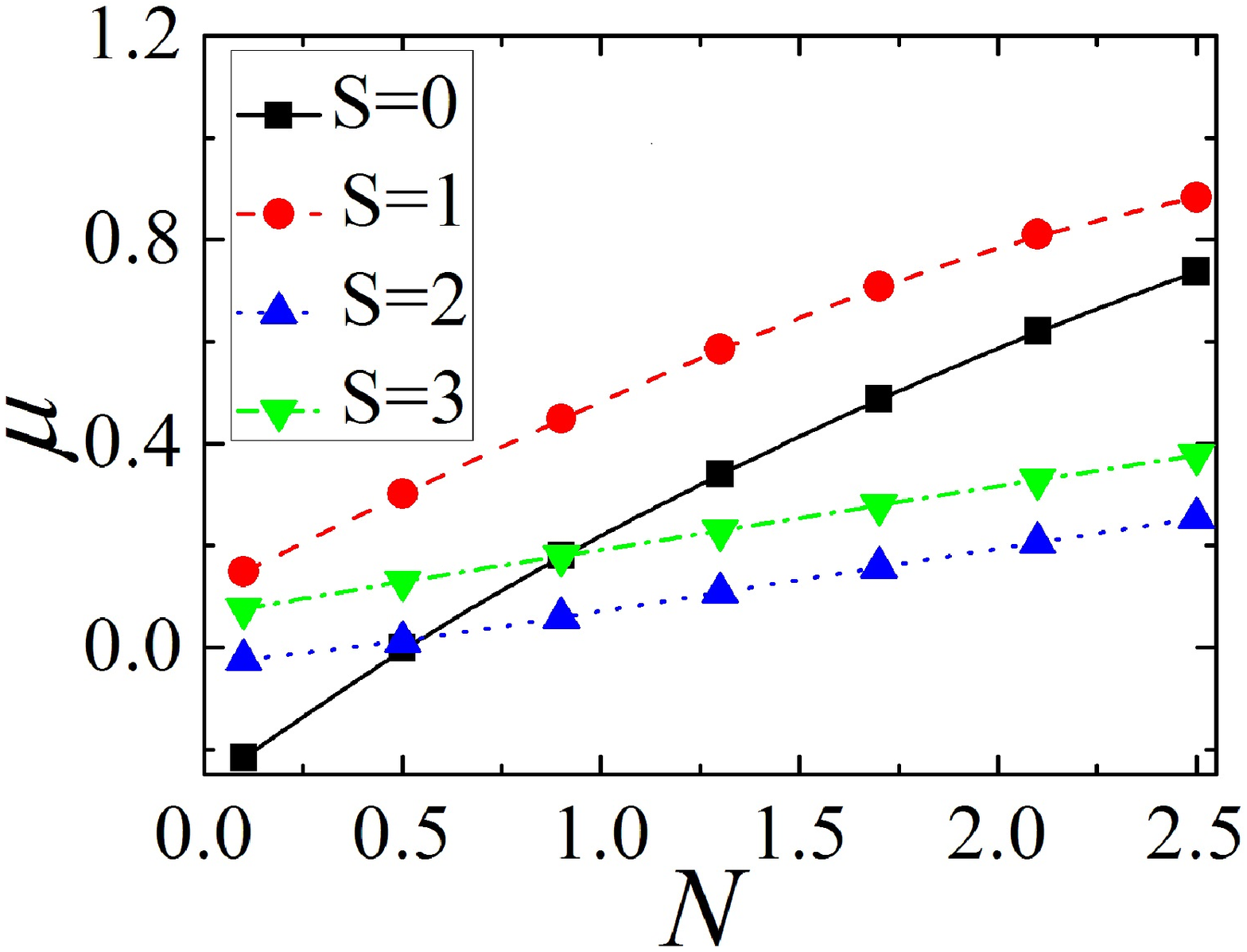}} %
\caption{(Color online) (a) The energy, defined by Eqs. (\protect\ref{E})-(%
\protect\ref{EDDI}), of the gap solitons with different vorticities $S$, and
the radial location of their main density peaks. Yellow stripes denote the
respective potential troughs. The norm of all the solitons presented in this
panel is $N=1.3$. (b) Chemical potential, $\protect\mu $, for the
gap-soliton families with different values of $S$, versus the norm, $N$, at $%
V_{0}=1$. This panel demonstrates that all the families satisfy the anti-
Vakhitov-Kolokolov criterion, $\mathrm{d}\protect\mu /\mathrm{d}N>0$, which
is a necessary condition for the stability of solitons supported by a
repusive nonlinearity (see the main text). }
\label{muVortex}
\end{figure}

Systematic numerical results, collected in Figs. \ref{abspsi} and \ref%
{muVortex}(a), demonstrates that the GSs with $S=0$ and $1$ are trapped in
the trough with $n=1$, GSs with $S=2$ and $3$ choose $n=2$, ones $S=4$ and $5
$ choose $n=3$, and so on, obeying an empiric relation%
\begin{equation}
n_{\mathrm{main~peak}}=1+\left[ S/2\right] ,  \label{main}
\end{equation}%
where $\left[ {}\right] $ stands for the integer part. Figure \ref{muVortex}%
(a) also shows that the vortex modes with odd $S$ have their energy
decreasing with the growth of $S$, while the energy of ones with even $S$
exhibit a very slow increase of the energy, starting from $S=2$. Further,
the mode with odd $S$ has its energy always higher than its counterpart with
even vorticity, $S-1$, sharing the same position of the density maximum.

The linear dependence in Eq. (\ref{main}) for large $S$ can be explained in
a qualitative form. Indeed, the strongest dependence of the energy of
ring-shaped solitons on the ring's radius, $R\approx \pi n$ for large $n$
[see Eq. (\ref{min})], is provided by the second term in Eq. (\ref{K}), $%
E_{S}~\approx \left( S/R\right) ^{2}N$ [for comparison, the DDI energy,
defined as per Eq. (\ref{EDDI}), can be estimated as $\sim \kappa N^{2}/R$].
On the other hand, the energy term which provides for the trapping of the
ring-shaped mode in the annular trap, is estimated as $E_{\mathrm{trap}%
}\approx 2V_{0}N$. Then , the balance of the two terms predicts $R\sim S/%
\sqrt{V_{0}}$.

The dependence of the chemical potential, $\mu $, on norm $N$, which is
displayed in Fig. \ref{muVortex}(b) for the GS families with $S=0,1,2,3$,
demonstrates that they obey the \textit{anti-Vakhitov-Kolokolov criterion}, $%
\mathrm{d}\mu /\mathrm{d}N>0$, which is a necessary condition for stability
of solitons supported by any kind of repulsive nonlinearity \cite{VK1,VK2}.
The same is true for still larger values of $S$, up to $S=11$ (largest $S$
considered in the present work).

\subsection{The radial lattice with $\protect\delta =\protect\pi $
(potential minimum at the center)}

\subsubsection{Fundamental gap solitons ($S=0$)}

\begin{figure*}[t]
\subfigure[]{\includegraphics[width=0.55\columnwidth]{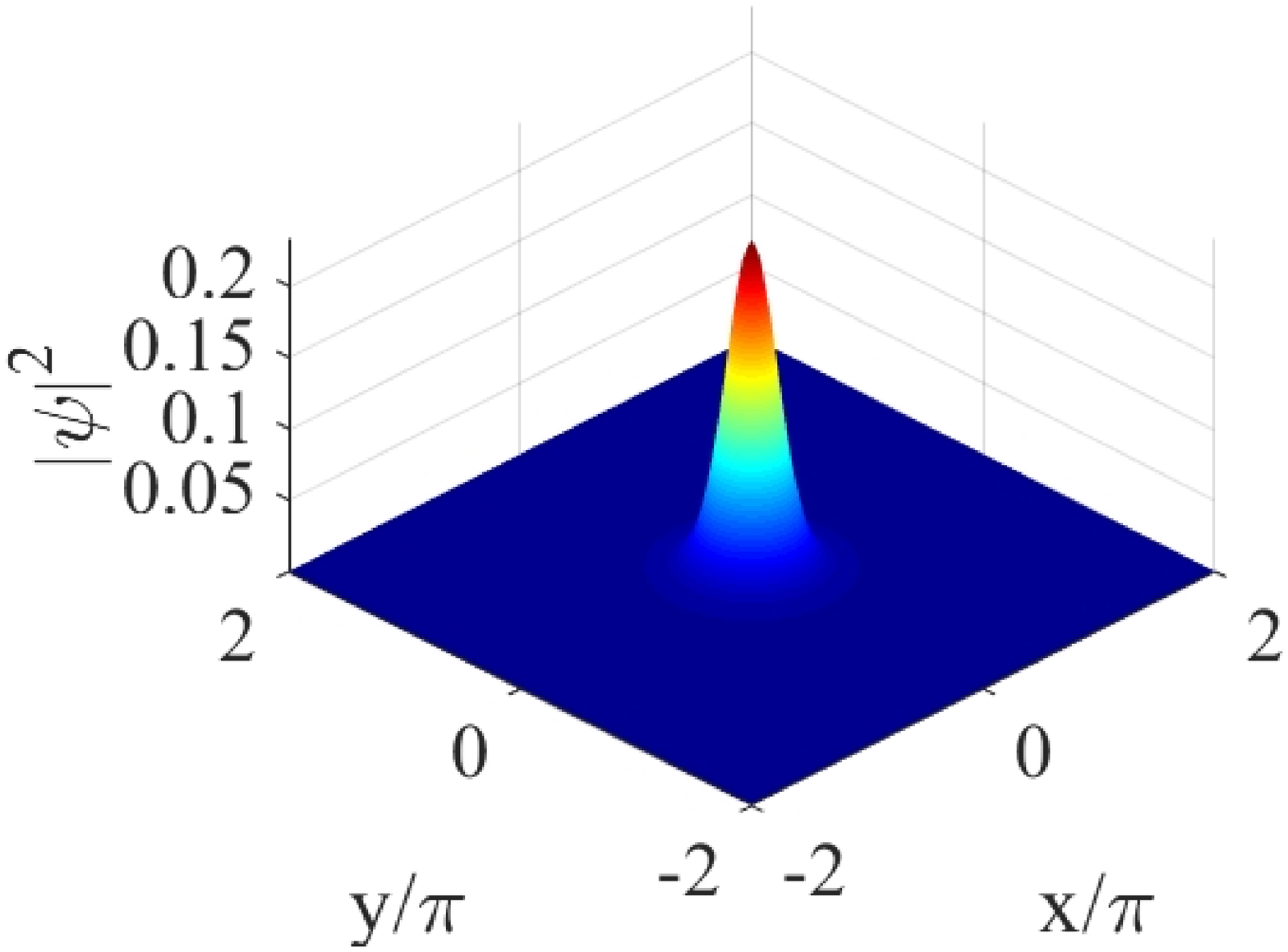}} %
\subfigure[]{\includegraphics[width=0.55\columnwidth]{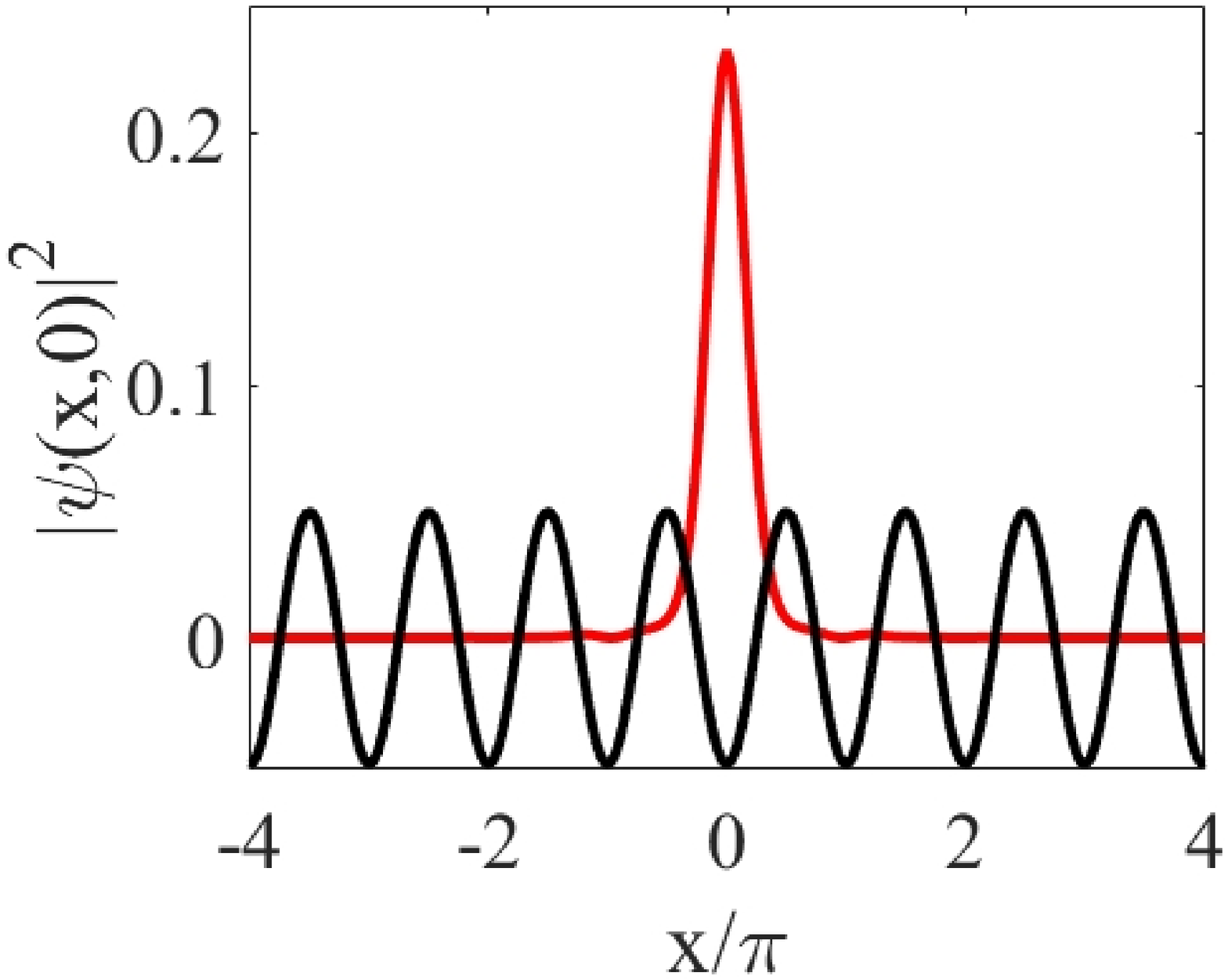}} %
\subfigure[]{\includegraphics[width=0.55\columnwidth]{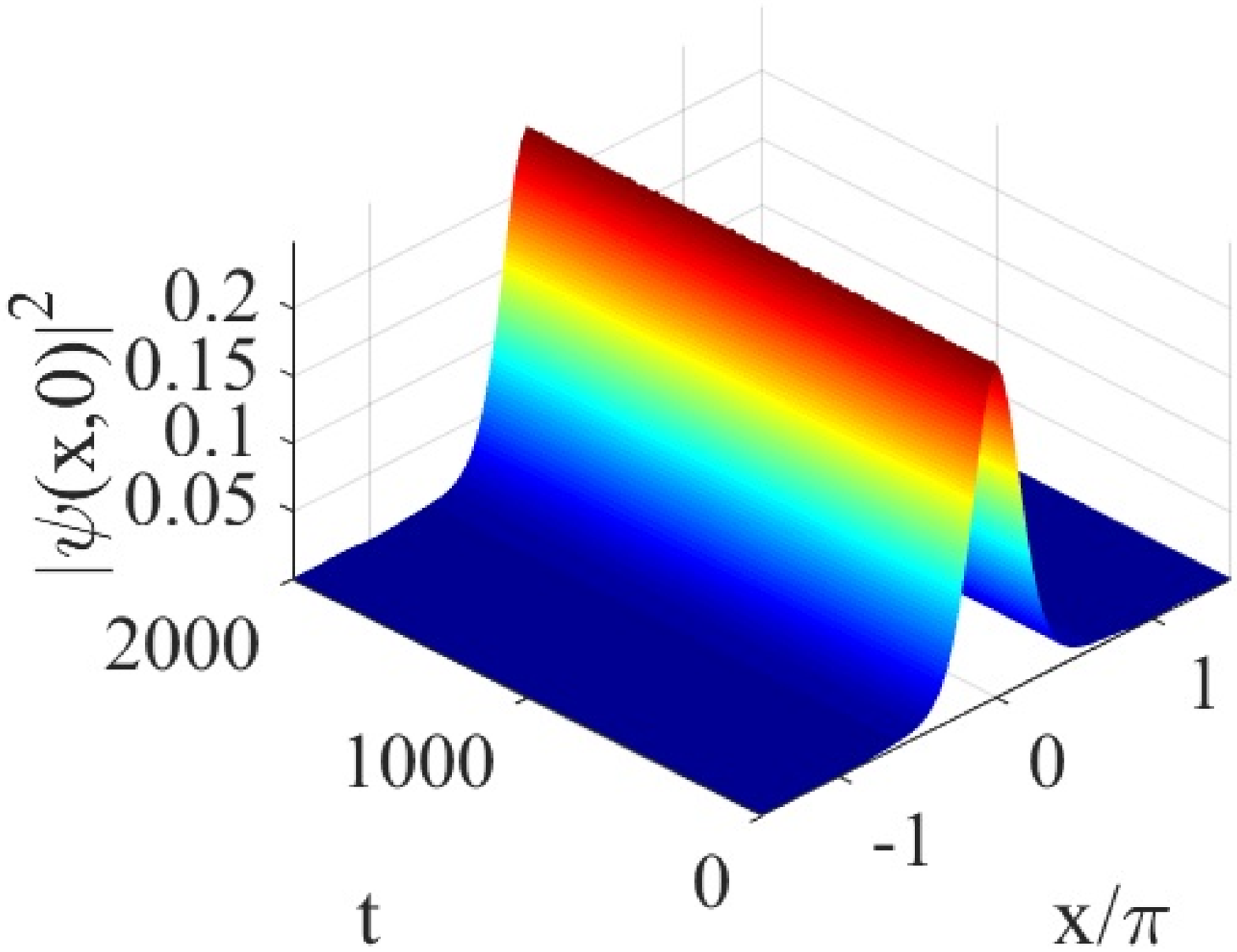}}
\caption{(color online). A fundamental ($S=0$) gap soliton trapped in the
center of the radial lattice potential (\protect\ref{ring}) with $\protect%
\delta =\protect\pi $. (a) The density profile of the soliton $|\protect\psi %
(x,y)|^{2}$. (b) Its cross-section, $|\protect\psi (x,0)|^{2}$. (c) The
cross-section of the simulated evolution of $|\protect\psi (x,0)|^{2}$,
which demonstrates stability of the soliton. Its norm is $N=0.5$, and the
strength of the potential is $V_{0}=2$ in this case. }
\label{Fundsoliton}
\end{figure*}

Radial potential (\ref{ring})\ with $\delta =\pi $ gives rise to a set of
potential minima%
\begin{equation}
r_{\min }=\pi n,~n=0,1,2,3,...,  \label{min2}
\end{equation}%
cf. Eq. (\ref{min}). In this case, is natural to expect the existence of
stable fundamental GSs with a density peak at $r=0$, which corresponds to
the zeroth minimum, in terms of Eq. (\ref{min2}). This expectation is borne
out by numerical results, see a typical example in Fig. \ref{Fundsoliton}.

\begin{figure*}[t]
\subfigure[]{\includegraphics[width=0.49\columnwidth]{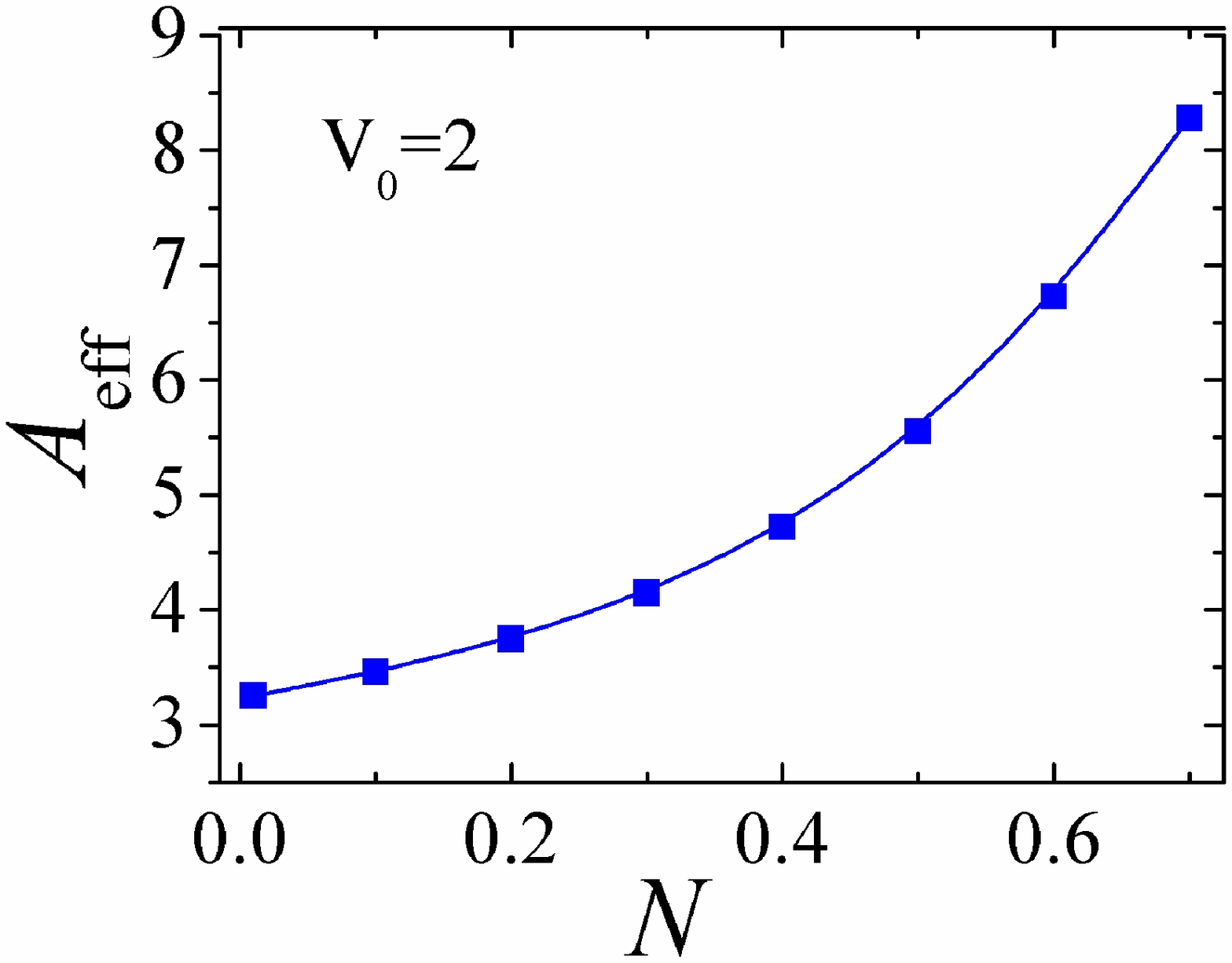}} %
\subfigure[]{\includegraphics[width=0.49\columnwidth]{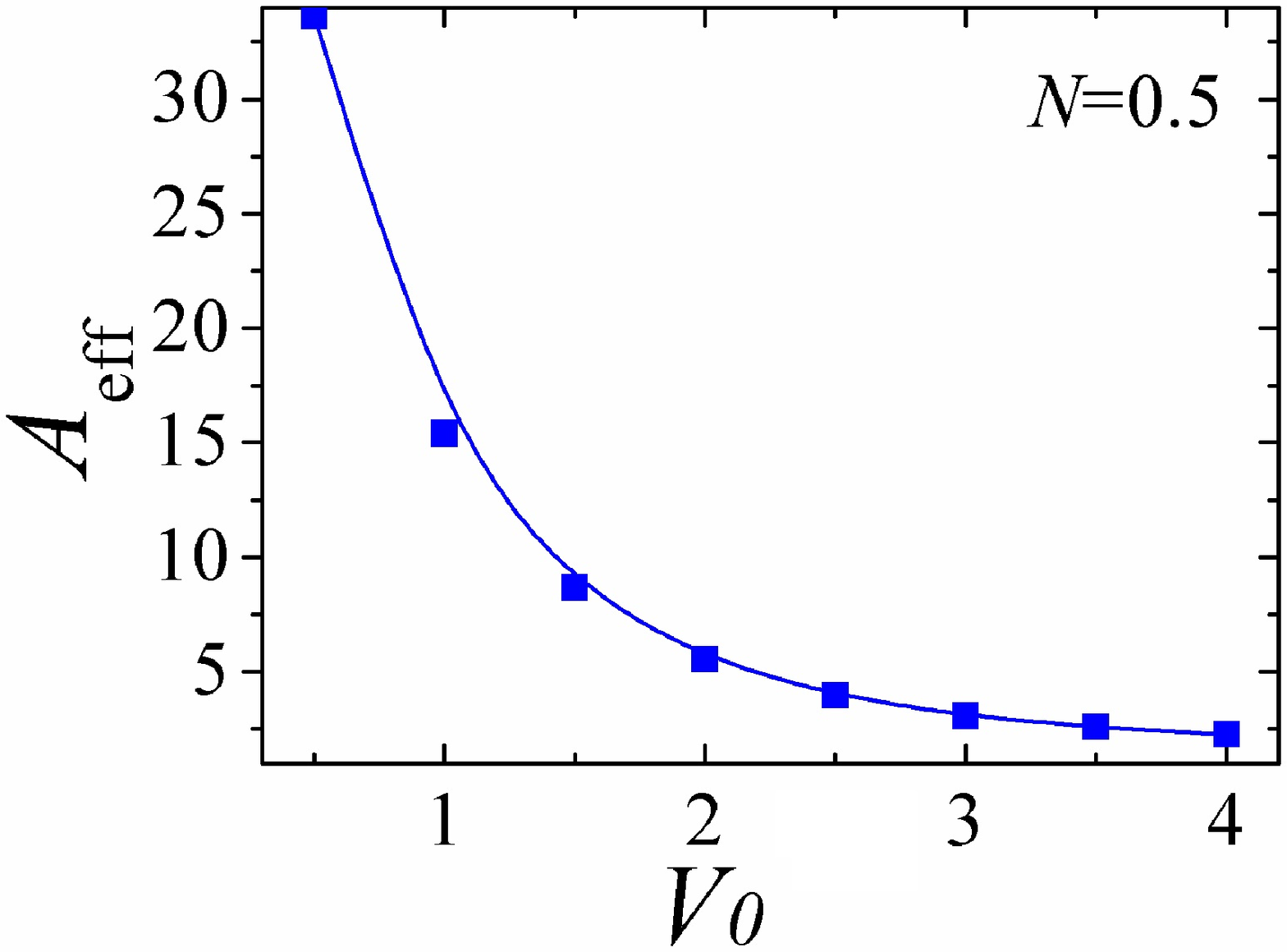}} %
\subfigure[]{\includegraphics[width=0.49\columnwidth]{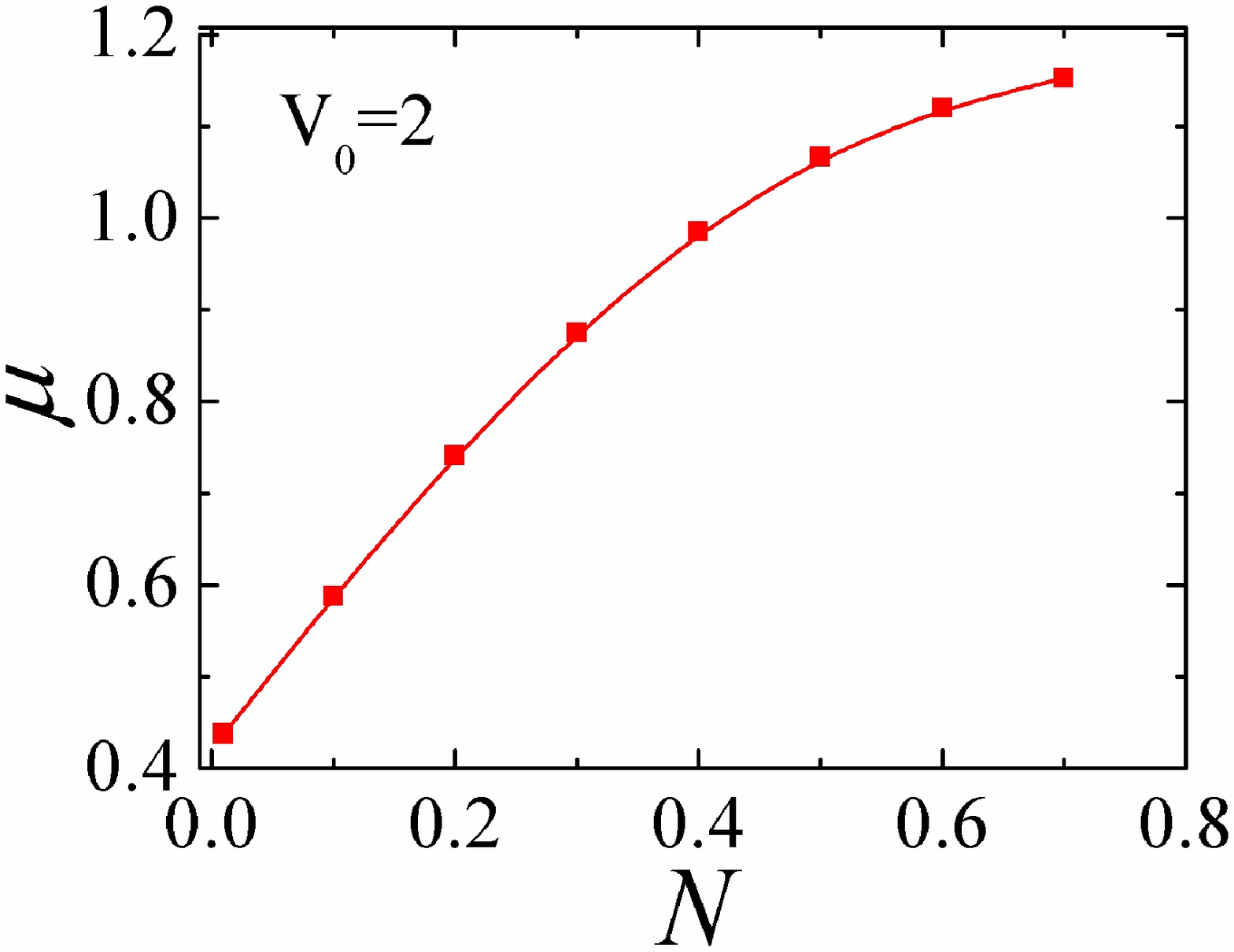}} %
\subfigure[]{\includegraphics[width=0.49\columnwidth]{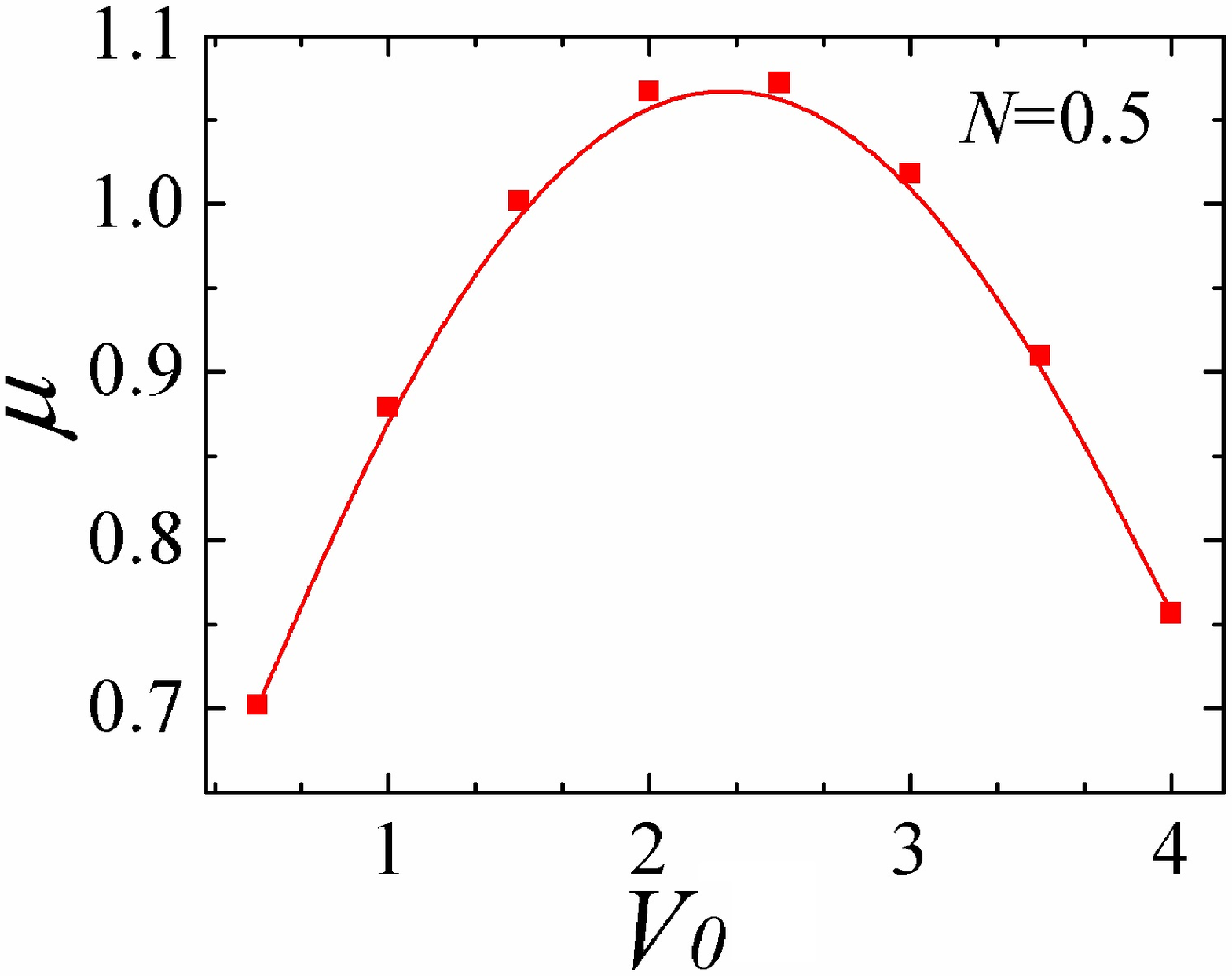}}
\caption{(Color online) (a,b) $A_{\mathrm{eff}}$ for the fundamental gap
soliton ($S=0$), trapped in the central well of radial potential with $%
\protect\delta =\protect\pi $ [$n=0$ in Eq. (\protect\ref{min2})], versus $N$
and $V_{0}$, respectively. In panel (a) $V_{0}=2$, and in (b) $N=0.5$. (c) $%
\protect\mu $ versus $N$ with $V_{0}=2$. (d) $\protect\mu $ versus $V_{0}$
with $N=0.5$. Other parameters are $\protect\kappa =1$ and $\protect\epsilon %
=0.5$.}
\label{Aeff}
\end{figure*}

To characterize the family of the fundamental GSs, we define its effective
area as
\begin{equation}
A_{\mathrm{eff}}={\frac{\left( \int |\psi (x,y)|^{2}\mathrm{d}x\mathrm{d}%
y\right) ^{2}}{\int |\psi (x,y)|^{4}\mathrm{d}x\mathrm{d}y}}.
\end{equation}%
Along with the chemical potential, $\mu $, it is shown, as a function of the
norm, $N$, and the potential depth, $V_{0}$, in Fig. \ref{Aeff}. In
particular, Fig. \ref{Aeff}(a,b) show that, quite naturally, the size of the
fundamental GS increases when the self-repulsive DDI becomes stronger, but
decreases with the growth of the trapping potential. Further, Fig. \ref{Aeff}%
(c) corroborates that these solitons satisfy the anti-Vakhitov-Kolokolov
criterion. In agreement with this finding, the family of the fundamental GSs
is entirely stable.

Finally, the transition from $\mathrm{d}\mu /\mathrm{d}V_{0}>0$ in the
relatively shallow radial lattice, at $V_{0}<2.3$, to $\mathrm{d}\mu /%
\mathrm{d}V_{0}<0$ at $V_{0}>2.3$ implies that properties of the soliton
family are dominated by the nonlinear interaction in the former case, and by
the linear trapping potential in the latter one. Indeed, direct simulations
demonstrate that the trapped mode with $S=0$ \textquotedblleft almost
exists" in the deep potential with $V_{0}>2.3$ in the absence of the
nonlinear interaction ($\kappa =0$), exhibiting very slow decay.

\subsubsection{Ring-shaped and vortex solitons}

\begin{figure*}[t]
\subfigure[]{\includegraphics[width=0.55\columnwidth]{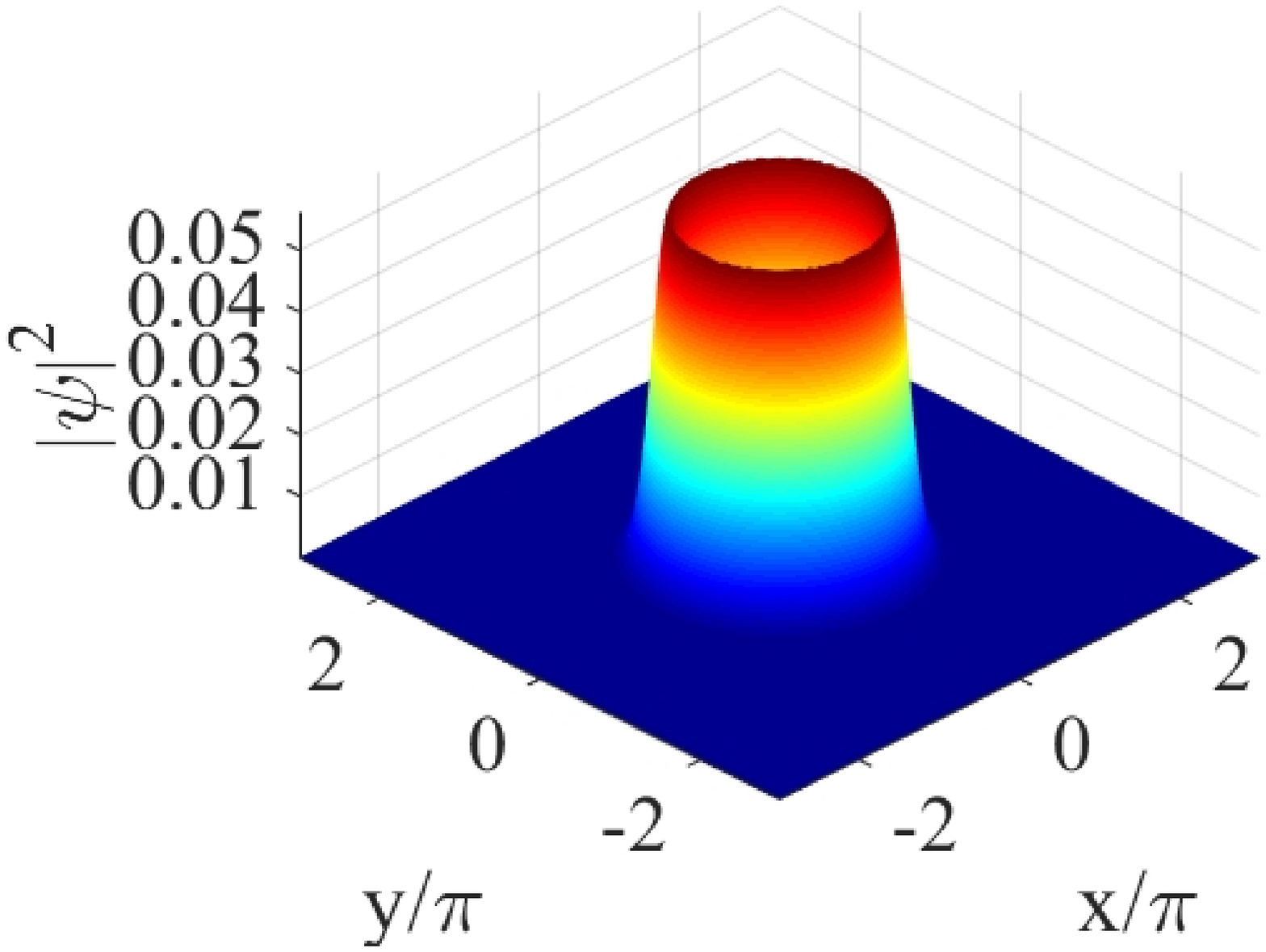}} %
\subfigure[]{\includegraphics[width=0.55\columnwidth]{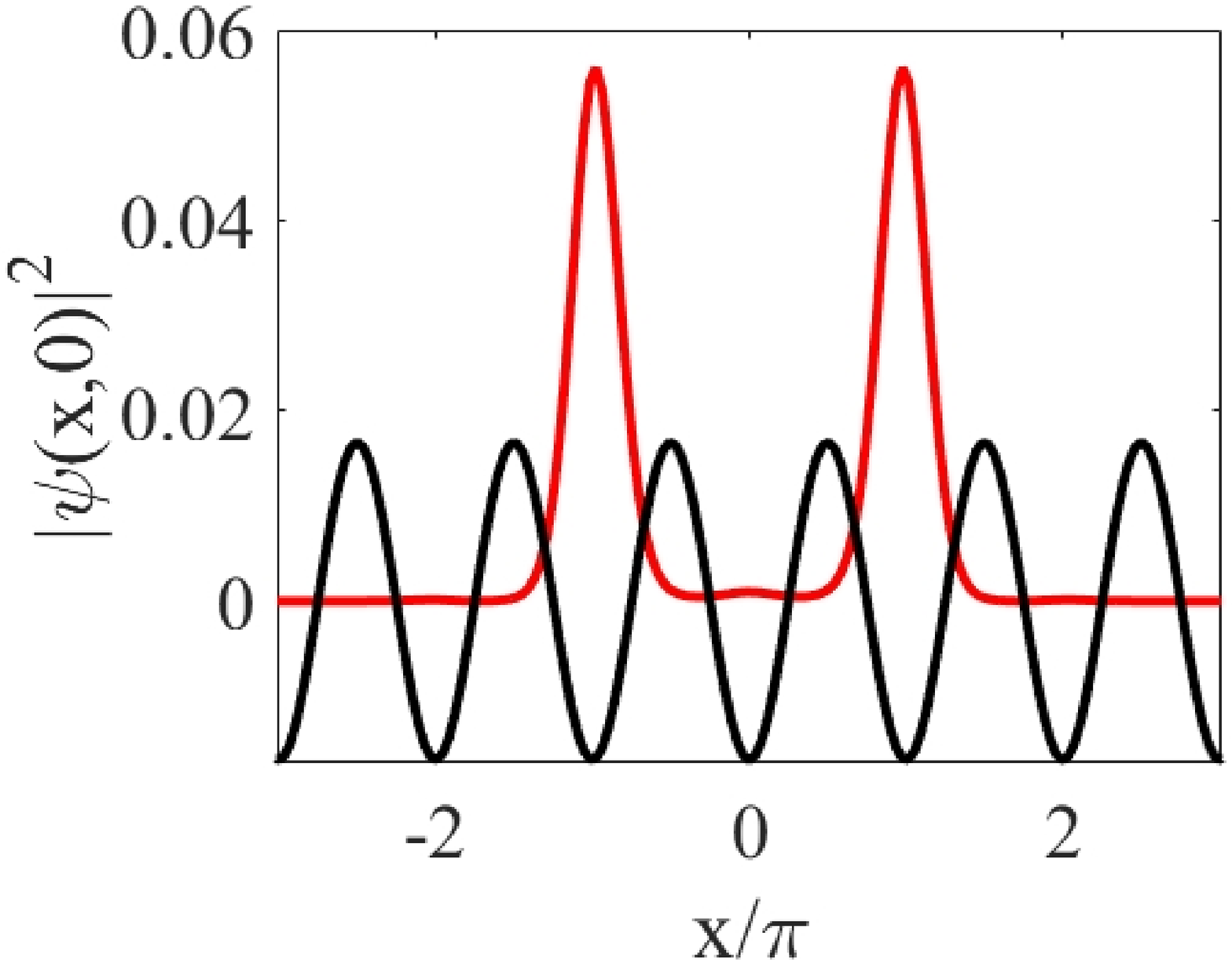}} %
\subfigure[]{\includegraphics[width=0.55\columnwidth]{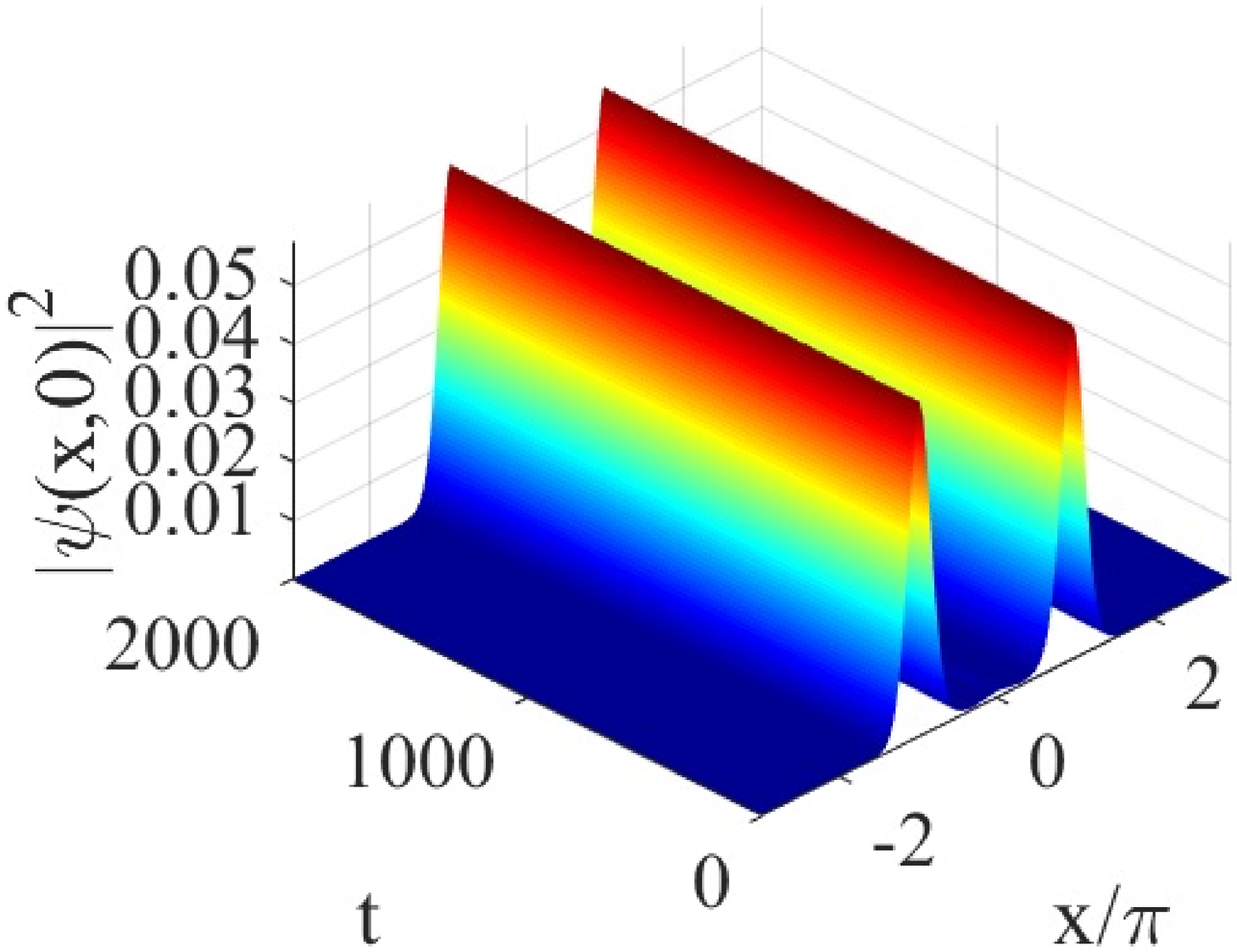}}
\caption{(Color online) A typical example of the fundamental ring soliton ($%
S=0$) trapped in the radial potential (\protect\ref{ring})\ with $\protect%
\delta =\protect\pi $. (a) The density profile of this gap soliton, $|%
\protect\psi (x,y)|^{2}$. (b) Its cross-section, $|\protect\psi (x,0)|^{2}$,
which clearly shows that it is trapped in the annular trough with $n=1$, see
Eq. (\protect\ref{min2}). (c) Cross-section of the simulated evolution,
which confirms the stability of the soliton. The norm of the soliton is $%
N=1.3$, and the half-depth of the radial potential is $V_{0}=2$. }
\label{anotherringsoliton}
\end{figure*}

\begin{figure*}[t]
\subfigure[]{\includegraphics[width=0.49\columnwidth]{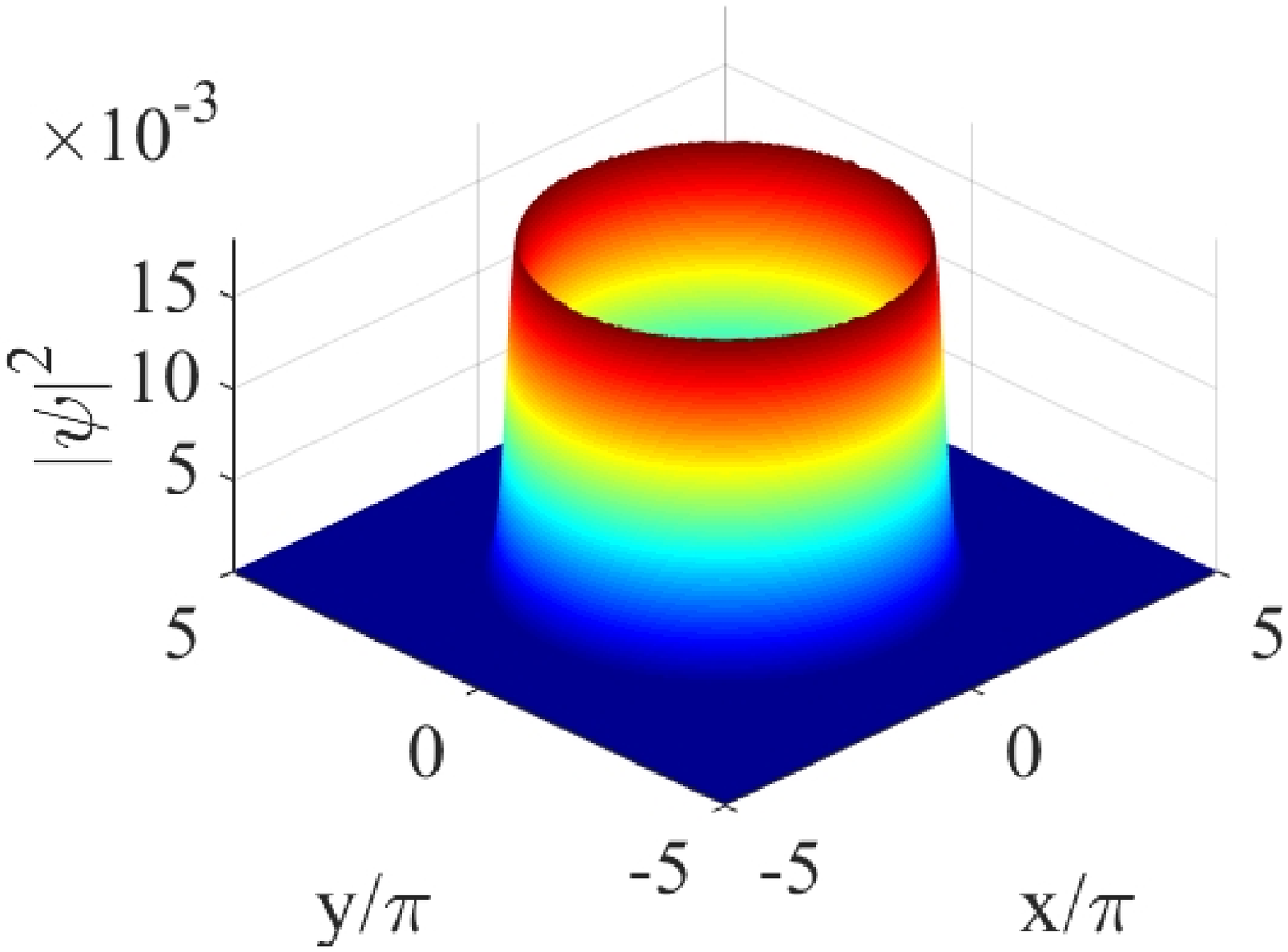}} %
\subfigure[]{\includegraphics[width=0.49\columnwidth]{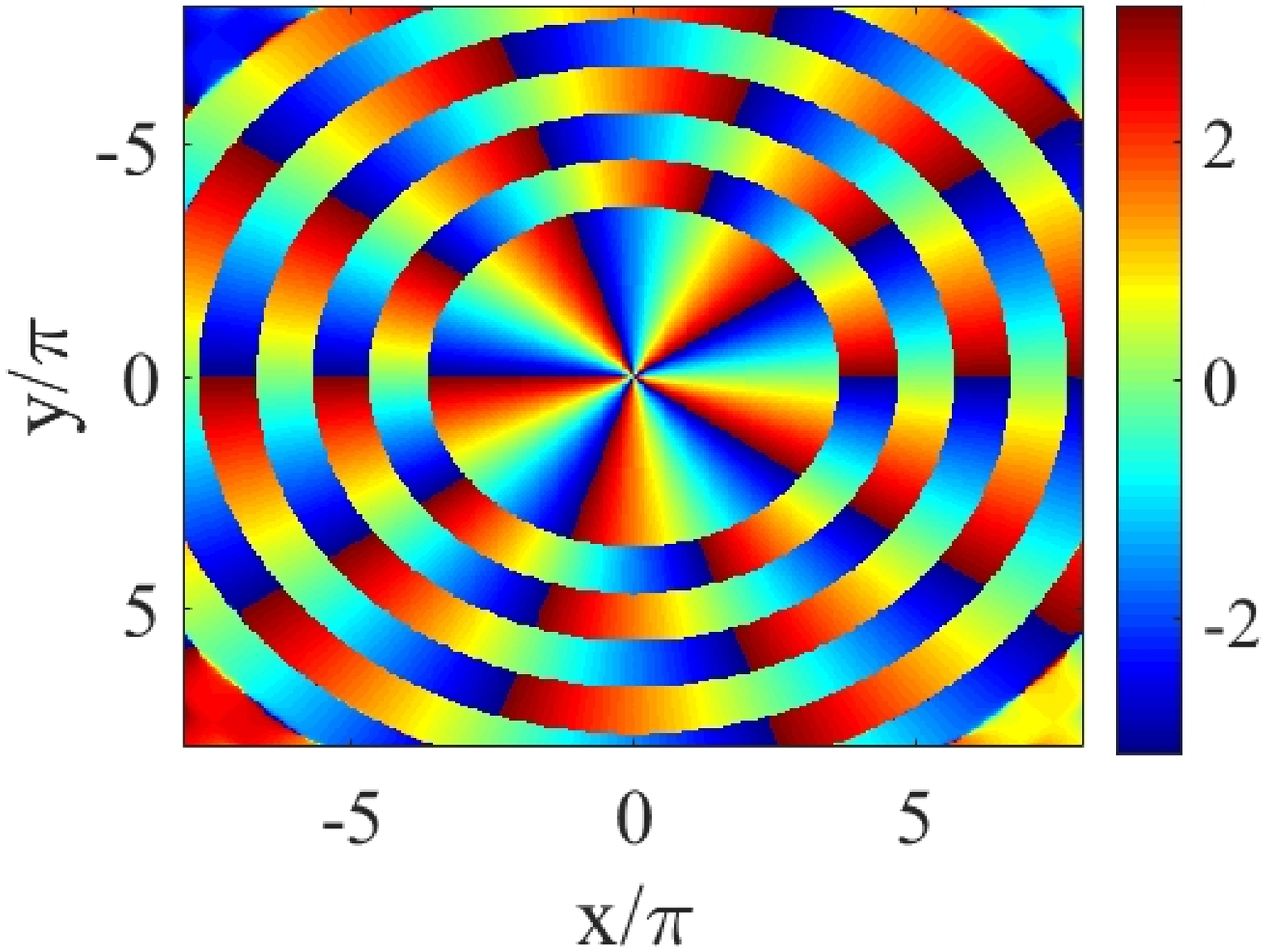}} %
\subfigure[]{\includegraphics[width=0.49\columnwidth]{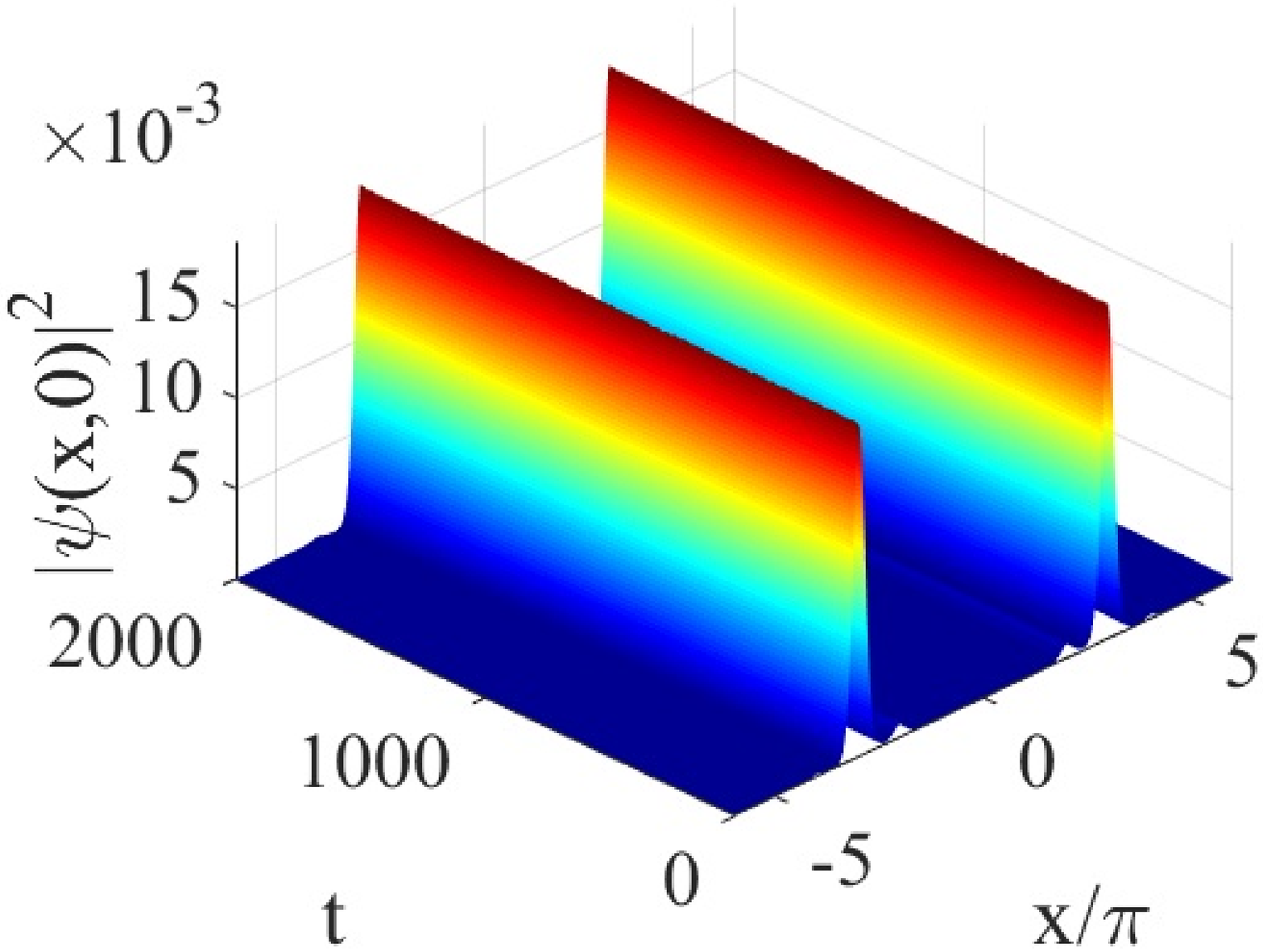}}%
\subfigure[]{\includegraphics[width=0.49\columnwidth]{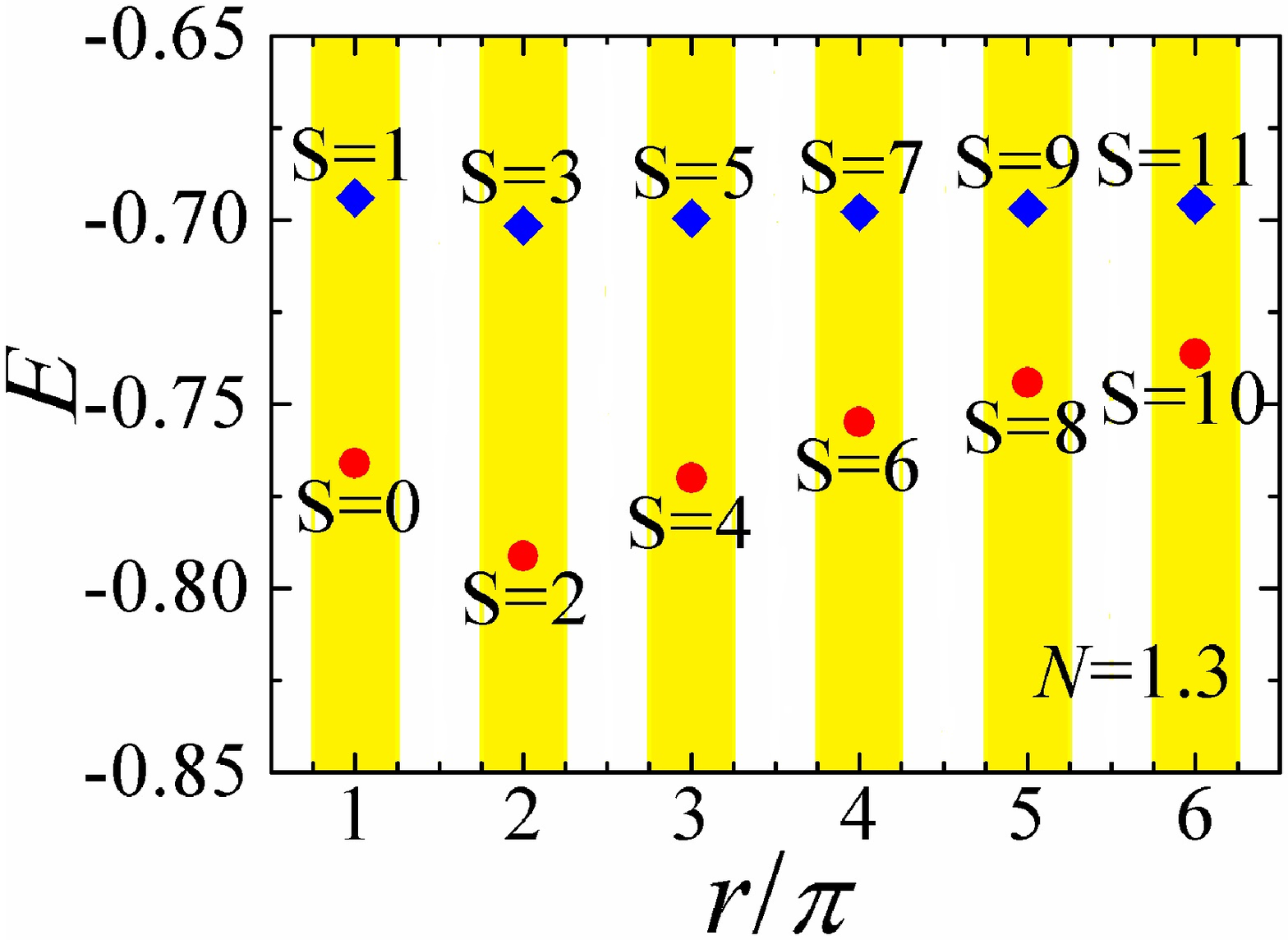}}
\caption{(Color online) An example of a stable vortex gap soliton with a
high topological charge, $S=5$, trapped in the third annular trough of the
radial potential (\protect\ref{ring}) with $\protect\delta =\protect\pi $.
Parameters are $N=1.3$ and $V_{0}=2$. (a) and (b): The density and phase
profiles of the vortex soliton. (c) The cross-section of the simultated
evolution, which corroborates the stability of the vortex. (d) The same as
in Fig. \protect\ref{muVortex}(a), but for potential (\protect\ref{ring})
with $\protect\delta =\protect\pi $  (the energy of the soliton with $S=0$
trapped at the center of the lattice, $n=0$, is $E_{0}=1.943$, which is not
displayed in this panel, as it is much larger than the values presented
here). Other parameters are fixed as $N=1.3$, $V_{0}=2$.}
\label{deltapiVortex}
\end{figure*}

Besides the fundamental GSs trapped in the central potential well, the
radial lattice with $\delta =\pi $ supports stable ring-shaped GSs, trapped
in annular potential trough, also with $S=0$, as well as ring-shaped vortex
GSs. A typical example of the stable ring mode with $S=0$, placed in the
trough with $n=1$, is displayed in Fig. \ref{anotherringsoliton}. Further,
Fig. \ref{deltapiVortex} shows an example of a stable soliton with high
vorticity, $S=5$, which is trapped in the trough with $n=3$ [see Eq. (\ref%
{min2})].

Numerical results produce the same relation between the location of the
ring-shaped solitons and their vorticity which is identified above for the
radial potential with $\delta =0$, see Eq. (\ref{main}) (the same
explanation for the linear dependence on large values of $S$, as that
outlined above, is relevant in the present case too). Namely, the GSs with $%
S=0$ and $1$ are trapped in the trough with $n=1$ in Eq. (\ref{min2}), ones
with $S=2$ and $3$ are placed at $n=2$, the solitons with $S=4$ and $5$ are
trapped in the trough with $n=3$, etc. These results are summarized in Fig. %
\ref{deltapiVortex}(d). Similar to the case of $\delta =0$ [cf. Fig. \ref%
{muVortex}(a)], the energy of the trapped states with odd $S$ is higher than
the energy of their counterparts with even vorticity, $S-1$, with the same
position of the density maximum. A difference from the case of $\delta =0$
is that the modes with even $S$ feature increase of their energy with the
growth of $S$, starting from $S=2$, while the energy of the modes with
odd $S$ remain virtually constant.

\subsection{Stable coexistence of double and multiple solitons}

\begin{figure*}[t]
\subfigure[]{\includegraphics[width=0.49\columnwidth]{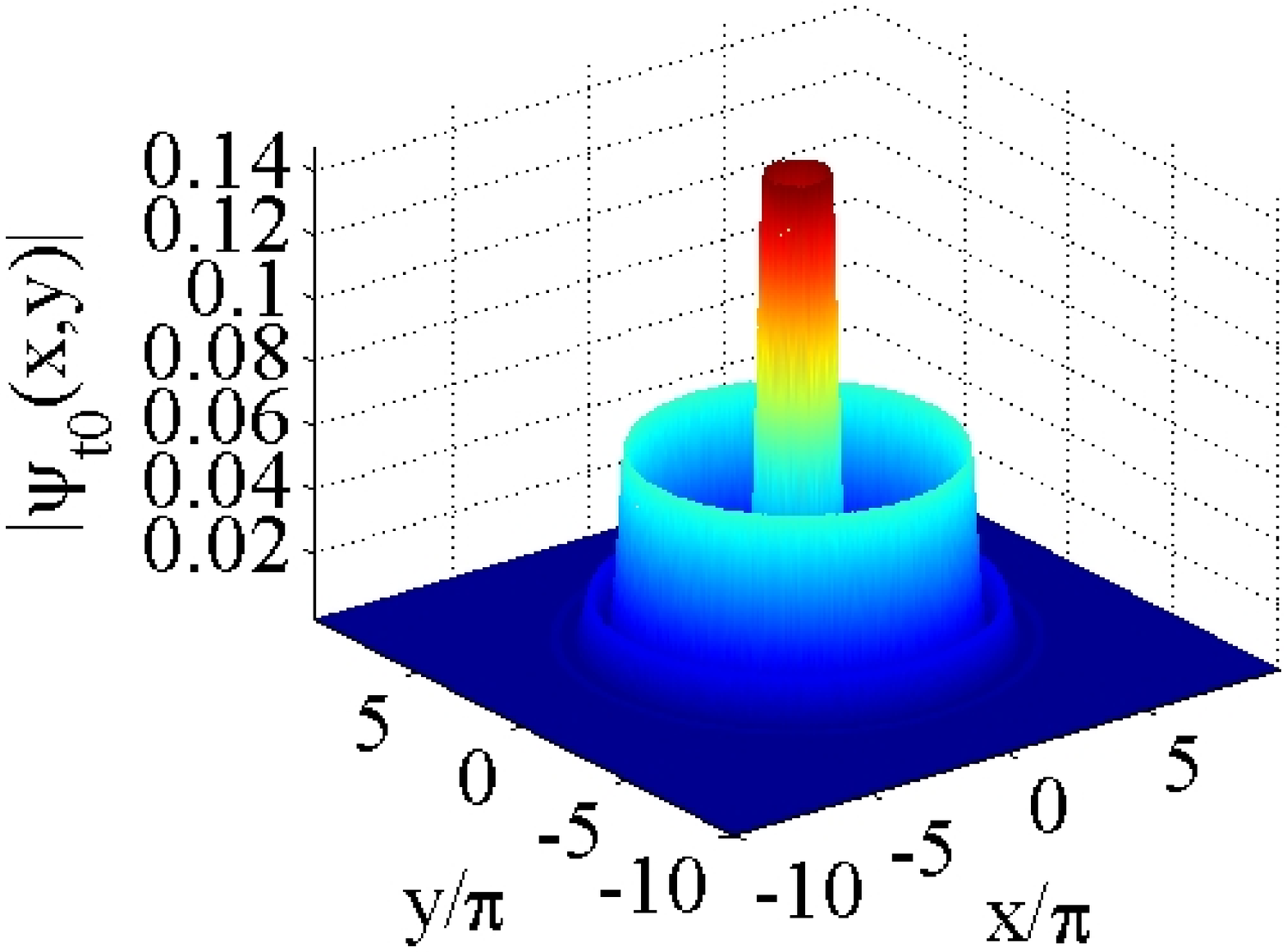}} %
\subfigure[]{\includegraphics[width=0.49\columnwidth]{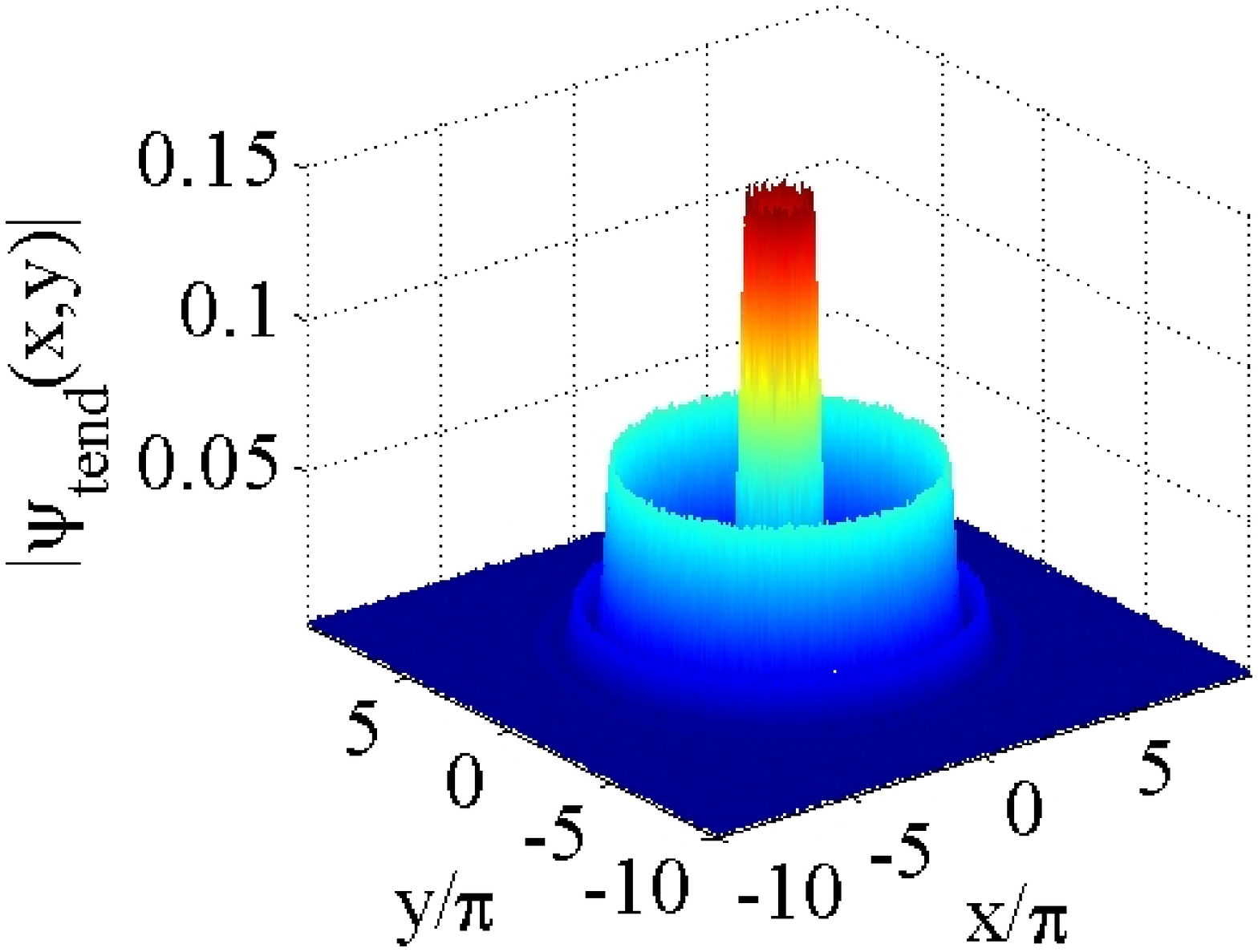}} %
\subfigure[]{\includegraphics[width=0.49\columnwidth]{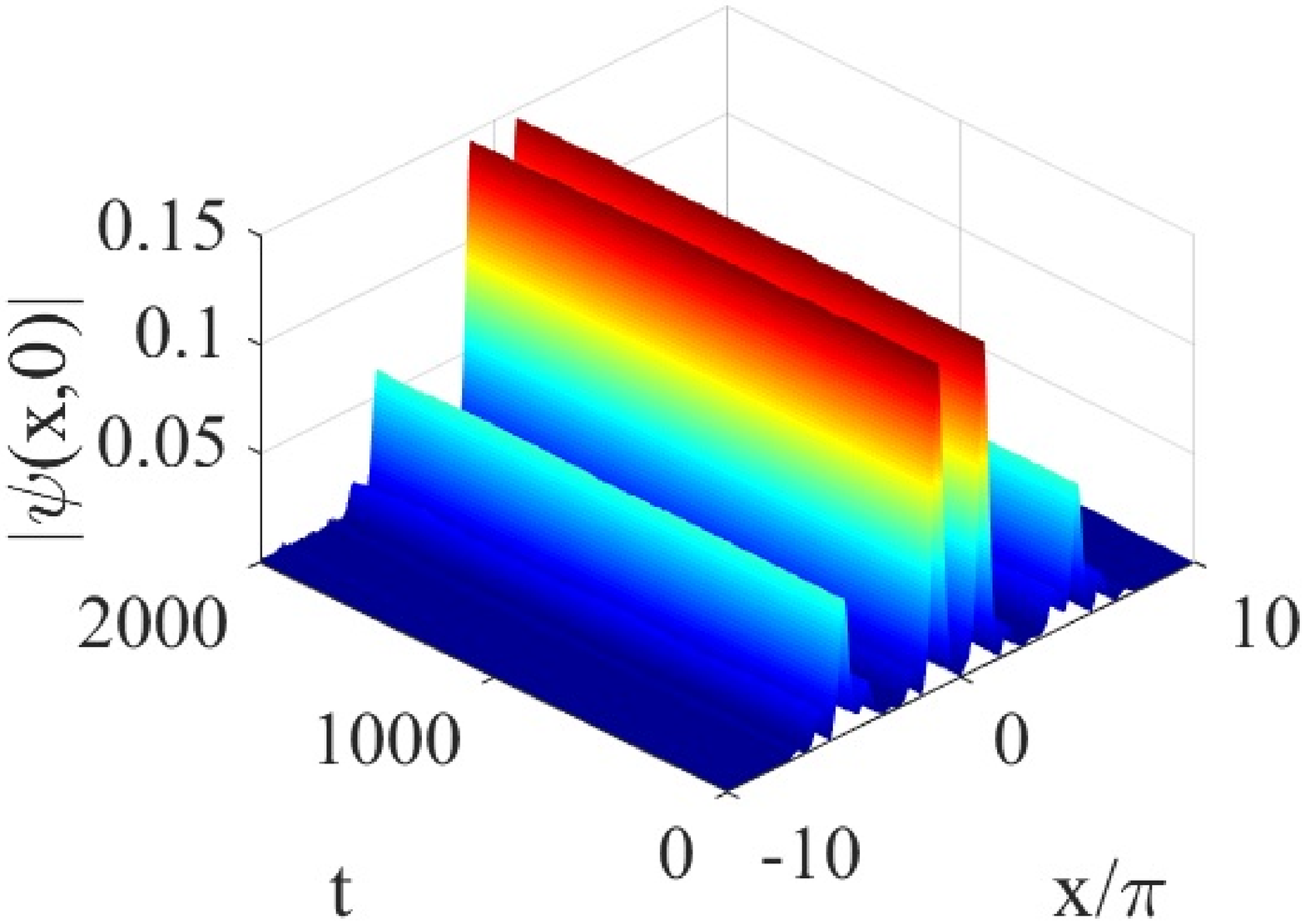}} %
\subfigure[]{\includegraphics[width=0.49\columnwidth]{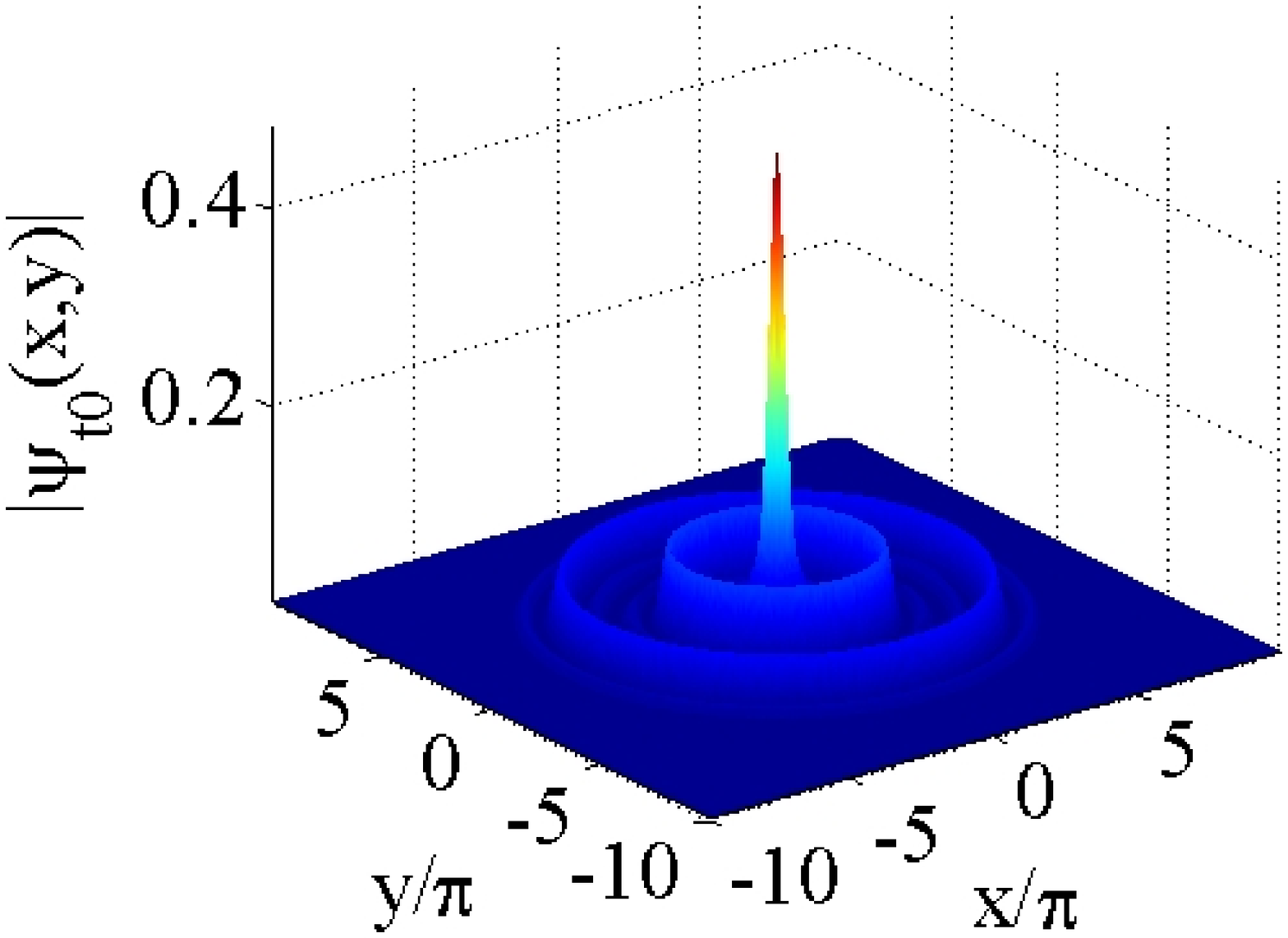}} %
\subfigure[]{\includegraphics[width=0.49\columnwidth]{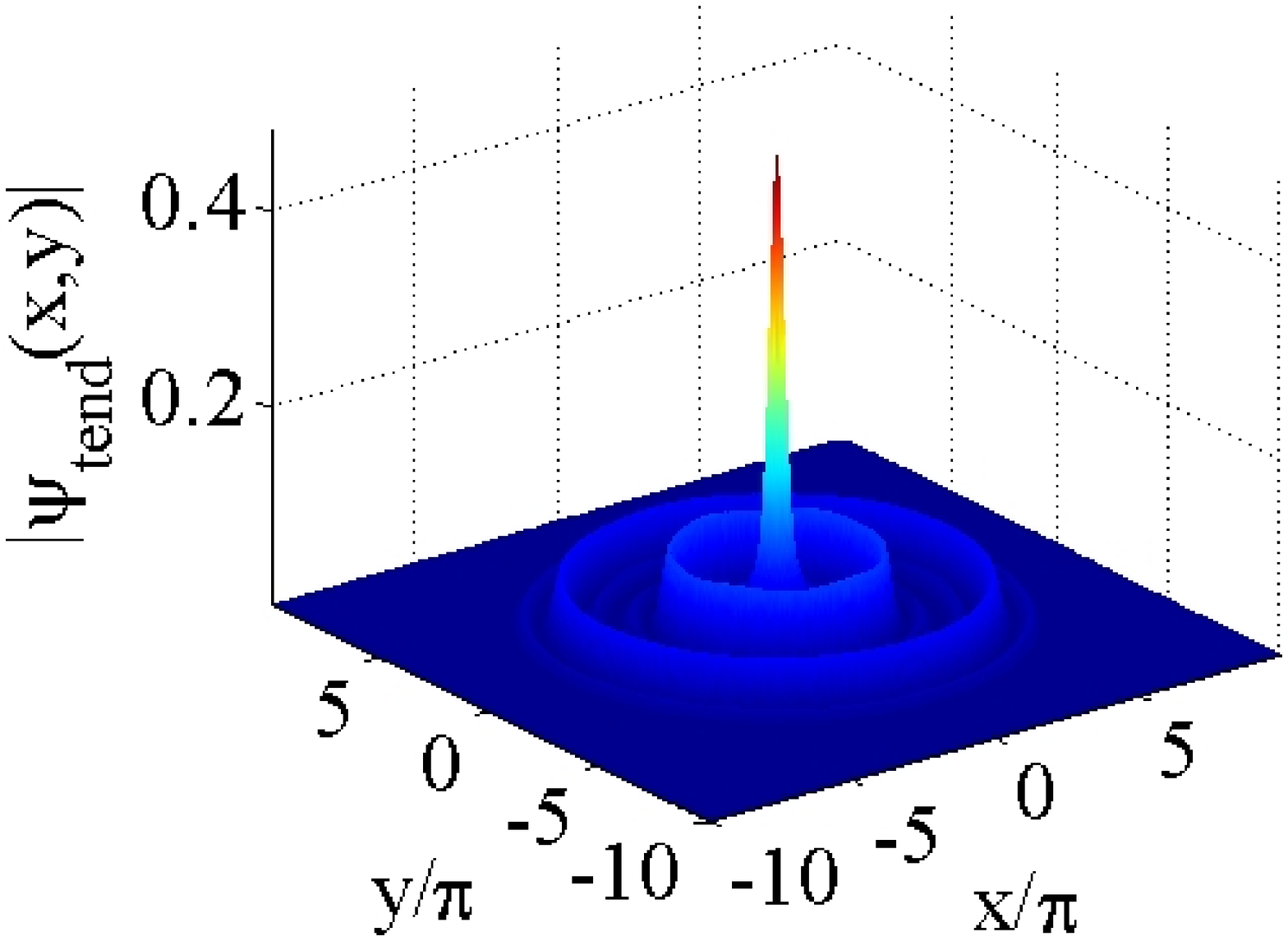}} %
\subfigure[]{\includegraphics[width=0.49\columnwidth]{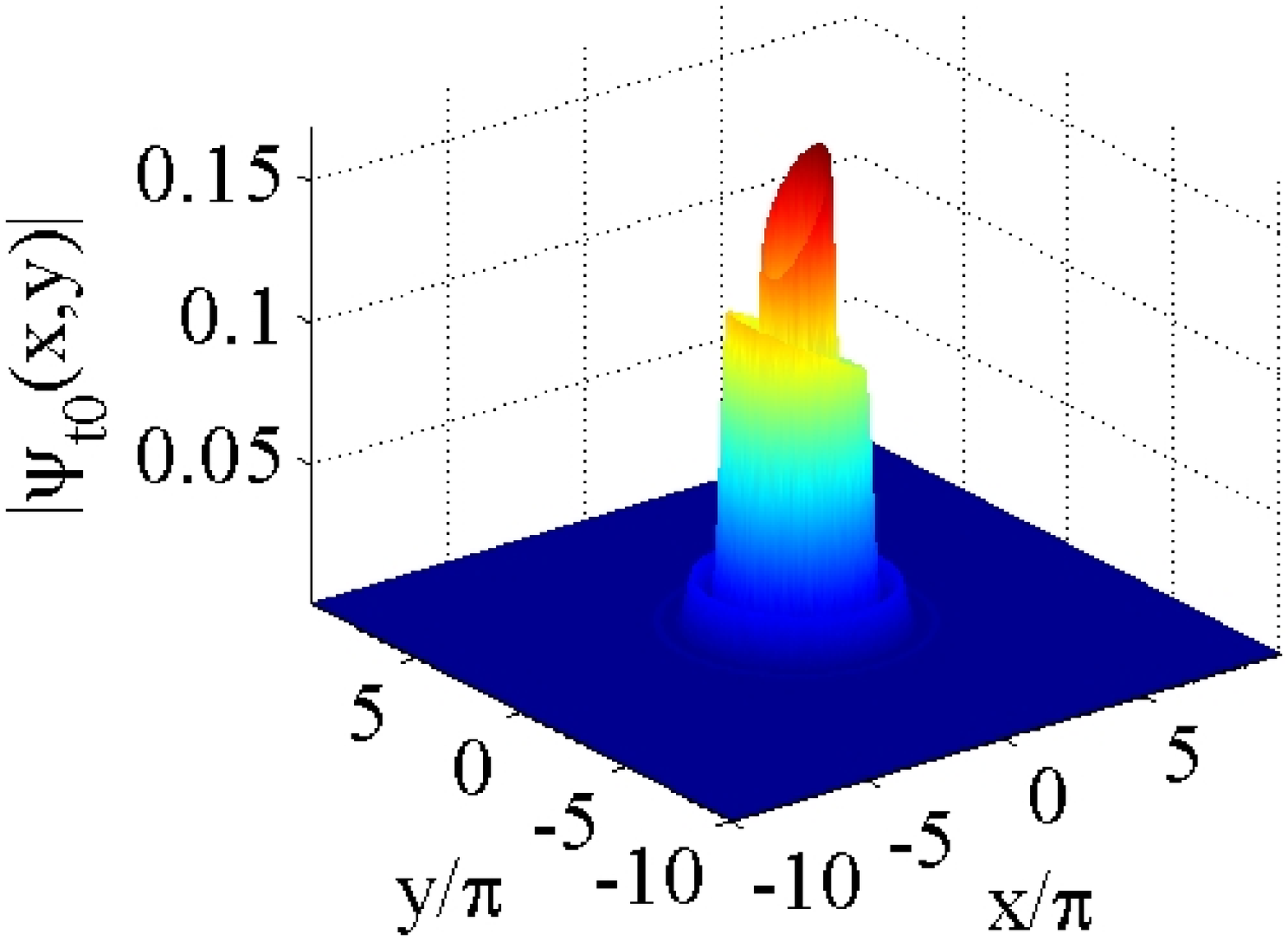}} %
\subfigure[]{\includegraphics[width=0.49\columnwidth]{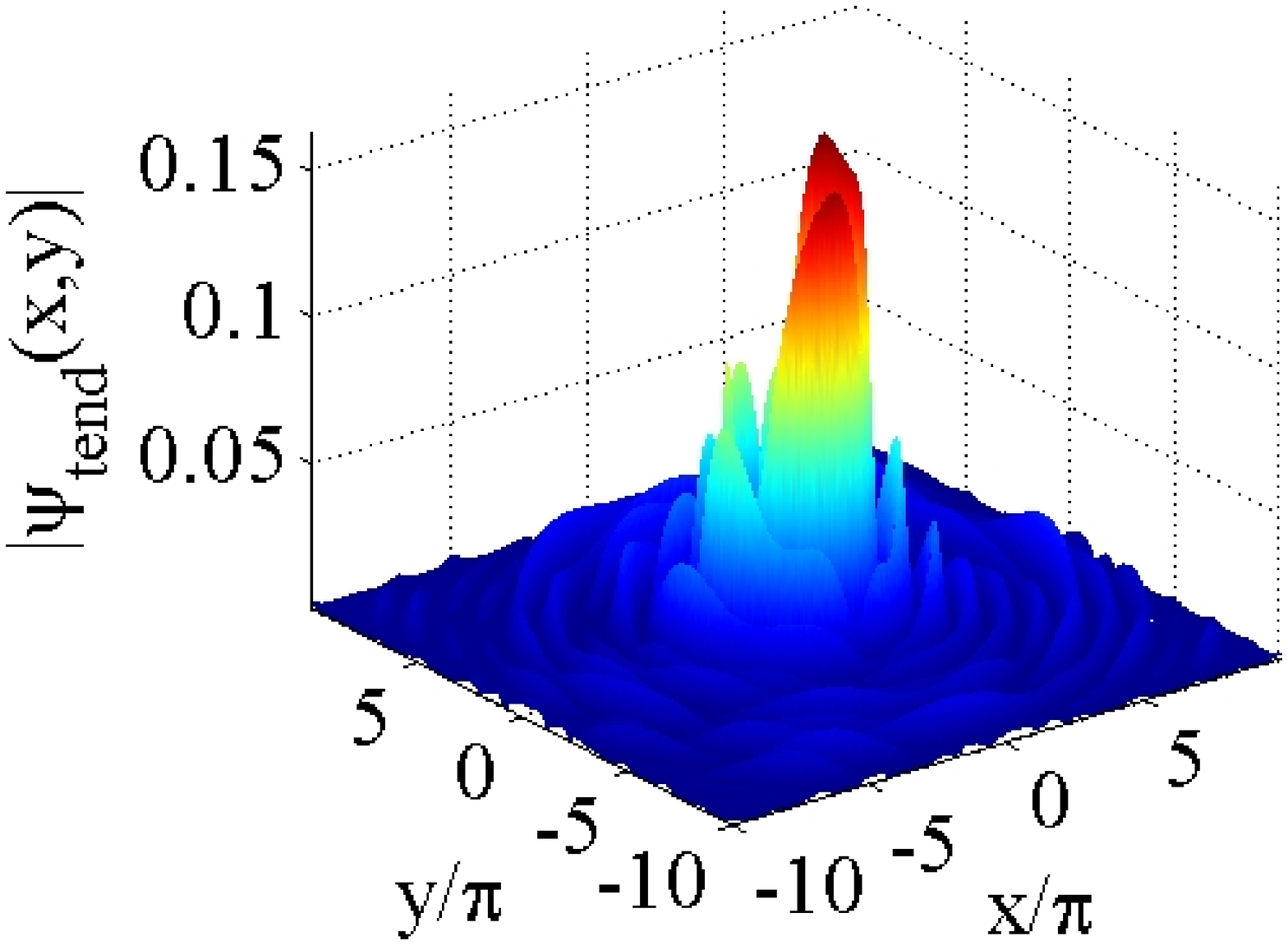}}
\caption{(Color online) (a) The absolute-value profile of the concentric
superposition of ring vortices with $S=1$ and $S=8$ (inner and outer rings,
respectively), used as an input for direct simulations, with potential (%
\protect\ref{ring}) that has a minimum at the center ($\protect\delta =%
\protect\pi $). (b) The output pattern of the simulation initiated by the
input in (a) at $t=2000$. (c) The cross-section along the $x$ axis,
illustrating the stable evolution of the concentric complex in the direct
simulations. (d) The same as in (a), but with the input formed by the
concentric superposition of three rings, with $S=0$ (placed at the the
center), $S=4$, and $S=10$. (e) The stable output pattern produced by the
evolution of the input from (d) at $t=2000$. (f) The same as in (a), but for
the input taken as a superposition of ring vortices with $S=1$ and $S=2$; in
this case, a conspicuous interference in observed in the input. (h) The
result of unstable evolution initiated by the input from (f). In all the
cases, parameters are $N=0.5$ and $V_{0}=2$.}
\label{Vortexsuperpose}
\end{figure*}

The existence of stable ring-shaped GSs with different topological charges,
located in different annular potential troughs, suggests that such modes
with different values of $S$ may have a chance to coexist in the system as
concentric modes, one embedded into the other. The coexistence of adjacent
layers with different topological charges implies that they must be
separated by zero-amplitude circular lines. Numerical results, produced by
direct simulations of inputs built as superpositions of two ring vortices
corroborate this conjecture, if the radial separation between the concentric
rings is large enough (i.e., the interaction between them is sufficiently
weak), as shown in Figs. \ref{Vortexsuperpose}(a-c). Due to the relation
between the radial location of the ring soliton and $S$ [as per Eq. (\ref%
{main})], the latter condition implies that the concentric rings must
pertain to sufficiently different values of $S$. Moreover, Figs. \ref%
{Vortexsuperpose}(d,e) demonstrate a similar result produced by the initial
superposition of three concentric GSs, under the same condition that they
are separated well enough. On the other hand, an input with conspicuous
overlap between the initial vortex rings gives rise to unstable evolution,
see Fig. \ref{Vortexsuperpose}(f,g).

\section{Conclusion}

We have elaborated a setting which makes it possible to readily stabilize
bright vortex solitons with arbitrarily high values of the topological
charge, $S$. The setting is realized as a two-dimensional dipolar BEC
trapped in an axisymmetric radially periodic potential, with dipole moments
of particles polarized perpendicular to the system's plane, which gives rise
to the isotropic repulsive DDI (dipole-dipole interaction). The radial
potentials with both the maximum and minimum at the center were considered.
The interplay of the radial lattice potential and repulsive interactions
creates families of stable annular GSs (gap solitons) with $S=0$ and $S\geq 1
$. Unlike the similar setting with contact repulsive interactions \cite%
{Bakhtiyor2006}, where the annular vortex GSs are (weakly) unstable, the
present system gives rise to GS families which are completely stable (at
least, up to $S=11$). The ring-shaped GSs have their main density peak
located in an annular potential trough whose number grows, for large $S$, as
$S/2$ [see Eq. (\ref{main})]. The linear growth of the vortex' radial
location with $S$ was qualitatively explained on the basis of the energy
considerations. Further, sets of concentric annular GSs with sufficiently
large radial separation between them, i.e., with essentially different
values of $S$, stably coexist in the present system. The optical angular
momentum becoming an important factor in modern information-processing
technologies \cite{OAM1,OAM2}, the setting analyzed in this work may find
application to the storage of data encoded in values of the vorticity.

A challenging extension of the work is to construct three-dimensional bright
GSs with embedded vorticity in the dipolar BEC, which will make it necessary
to combine the radial potential with the a term which periodically varies
along the transverse coordinate, $z$ [such as $\cos \left( qz\right) $], so
as to build an axially stacked version of the radial potential lattice, that
should feature a full three-dimensional bandgap in its spectrum.

\section*{Acknowledgments}

This work is supported by the National Natural Science Foundation of China
(Grant Nos. 11575063, 11204037, 61575041). B.A.M. appreciates a Ding Ying
visiting professorship provided by the South China Agricultural University
(Guangzhou). The work of this authors is supported, in part, by the joint
program in physics between NSF and Binational (US-Israel) Science Foundation
through project No. 2015616, and by the Israel Science Foundation through
grant No. 12876/17.

\bibliographystyle{plain}
\bibliography{apssamp}

\end{document}